\documentclass[aps,prd,twocolumn,superscriptaddress, nofootinbib, longbibliography, notitlepage]{revtex4-1}
\usepackage{epsf,epsfig,graphicx}
\usepackage{amsmath}
\usepackage{amssymb}
\usepackage{tikz}
\usetikzlibrary{arrows.meta}
\usetikzlibrary{calc}
\usepackage{subfigure}
\usepackage{multirow}
\usepackage{tabularx}

\usepackage{bbding,enumitem}

\usepackage{pifont} 

\makeatletter
\DeclareFontFamily{OMX}{MnSymbolE}{}
\DeclareSymbolFont{MnLargeSymbols}{OMX}{MnSymbolE}{m}{n}
\SetSymbolFont{MnLargeSymbols}{bold}{OMX}{MnSymbolE}{b}{n}
\DeclareFontShape{OMX}{MnSymbolE}{m}{n}{
    <-6>  MnSymbolE5
   <6-7>  MnSymbolE6
   <7-8>  MnSymbolE7
   <8-9>  MnSymbolE8
   <9-10> MnSymbolE9
  <10-12> MnSymbolE10
  <12->   MnSymbolE12
}{}
\DeclareFontShape{OMX}{MnSymbolE}{b}{n}{
    <-6>  MnSymbolE-Bold5
   <6-7>  MnSymbolE-Bold6
   <7-8>  MnSymbolE-Bold7
   <8-9>  MnSymbolE-Bold8
   <9-10> MnSymbolE-Bold9
  <10-12> MnSymbolE-Bold10
  <12->   MnSymbolE-Bold12
}{}

\let\llangle\@undefined
\let\rrangle\@undefined
\DeclareMathDelimiter{\llangle}{\mathopen}%
                     {MnLargeSymbols}{'164}{MnLargeSymbols}{'164}
\DeclareMathDelimiter{\rrangle}{\mathclose}%
                     {MnLargeSymbols}{'171}{MnLargeSymbols}{'171}
\makeatother

\RequirePackage[colorlinks,citecolor=red,urlcolor=red,linkcolor=red]{hyperref}

\def\edth{\;\raise1.0pt\hbox{$'$}\hskip-6pt\partial\;}
\def\baredth{\;\overline{\raise1.0pt\hbox{$'$}\hskip-6pt
\partial}\;}
\def\gsim{~\rlap{$>$}{\lower 1.0ex\hbox{$\sim$}}}

\newcommand{\be}{\begin{equation}}
\newcommand{\ee}{\end{equation}}
\newcommand{\bw}{\begin{widetext}}
\newcommand{\ew}{\end{widetext}}

\usepackage{xcolor}

\usepackage[normalem]{ulem}

\AtBeginDocument{}

\begin{document}

\title{Stochastic problems in pulsar timing}

\author{Reginald Christian Bernardo}
\email{reginald.christian.bernardo@aei.mpg.de}
\affiliation{Max Planck Institute for Gravitational Physics (Albert Einstein Institute), Hannover 30167, Germany}



\begin{abstract}
Langevin stochastic differential equations provide a dynamical description of pulsar timing noise and gravitational wave background (GWB) signals. They are also central to state space algorithms that have gained traction in pulsar timing array analysis due to their linear computational scaling with the number of observations. In this work, we utilize established methods in diffusion theory to derive analytical time-domain solutions (means, covariances, and probability density functions) to Langevin equations relevant to red noise and the GWB signal in pulsars. The solutions give direct physical insight on the dynamics of pulsar timing signals. As a canonical example, we show that the pulsar spin frequency modeled as an Ornstein-Uhlenbeck process is mathematically inconsistent with a stationary GWB signal when the timing residual is the direct observable. The nonstationarity can be partially dealt with by marginalizing over long time deterministic trends in the data. Then, we show that a random process based on an overdamped harmonic oscillator supports both a stationary spin frequency and phase residuals, consistent with a stationary GWB signal. We also turn our attention to a phenomenological model of a neutron star -- a two-component model with spin wandering -- that has been motivated to explain observed timing noise in radio pulsars. We derive analytical expressions for the means, covariances, and cross-covariances of the crust and superfluid rotational states driven by white noise. The associated constant deterministic torques are linked to the quadratic spin-down of pulsars. The solutions reveal the physical origin of nonstationarity in the residual model: the coexistence of damped and diffusive eigenmodes of the system.
\end{abstract}

\maketitle

\section{Introduction}
\label{sec:introduction}

Pulsars are the Universe's most precise clocks \cite{Hewish:1968bj, Pilkington:1968bk, Wallace:1977he, Taylor:1977wj}, with old, stable neutron stars spinning at frequencies of tens to hundreds of hertz and exhibiting fractional changes in their rotation rates at the level of one part in a trillion \cite{Lorimer:2001vd, Kaspi:2016jkv, Bassa:2017zpe, Fermi-LAT:2018jvx, Nieder:2020yqy, Clark:2025lai, Bagchi:2025hnw}. This rotational stability enables the prediction of pulse arrival times with exquisite precision, facilitating an interplay between pulsar astronomy and fundamental physics and astrophysics, including tests of strong gravity \cite{Manchester:2015mda, Kramer:2021jcw, Freire:2024adf}, constraints on dark matter \cite{Khmelnitsky:2013lxt, Porayko:2014rfa, Porayko:2018sfa, EuropeanPulsarTimingArray:2023egv}, and the detection and characterization of gravitational waves (GW) \cite{Hulse:1974eb, 1982ApJ...253..908T, Hulse:1994zz, Damour:2014tpa}. Pulsar timing arrays (PTAs) leverage this stability by analyzing the spatial correlation patterns in the mean \cite{Hellings:1983fr} and variance \cite{Roebber:2016jzl, Allen:2022dzg, Bernardo:2022xzl} of timing residuals, which are induced by compact binary sources \cite{Sazhin:1978myk} or a gravitational wave background (GWB) \cite{Detweiler:1979wn, Allen:1987bk, Sesana:2008mz}.

The timing residual, defined as the difference between observed and predicted pulse arrival times based on a deterministic timing model, is central to pulsar science's pursuit of GW detection. Unlike measurement errors, timing residuals encode valuable information about physical processes affecting pulsars and spacetime. A key challenge in PTA science is accurately modeling and interpreting the stochastic components of these residuals \cite{Bi:2023tib, Allen:2024uqs, Goncharov:2024htb, Bernardo:2024tde, Xue:2024qtx, Crisostomi:2025vue, Bernardo:2025dat, Allen:2025waa, Yu:2025dor}.

Intrinsic pulsar timing noise and the nanohertz GWB both manifest as time-correlated random processes with comparable time scales \cite{1982MNRAS.199..659M, 1985MNRAS.217..265M, vanHaasteren:2008yh, Coles:2011zs, vanHaasteren:2012hj, vanHaasteren:2014qva}. The internal physics of neutron stars drives spin irregularities, resulting in red-noise-like fluctuations \cite{Melatos:2013rca, Haskell:2015jra}, while a GWB, generated by numerous unresolved sources, introduces both temporal and spatial correlations. These signals are typically modeled as Gaussian processes, characterized by their spectral properties \cite{Phinney:2001di, Goncharov:2019mei, Galikyan:2026zjl} and, for GWs, by their spatial correlation patterns \cite{Renzini:2022alw, Bernardo:2024bdc}.
Complementing the Gaussian process approach, this work adopts a physically transparent perspective by viewing pulsar timing observables -- including rotational state, timing, and phase residuals -- as manifestations of random walk dynamics. This perspective is grounded in the established equivalence between Gaussian processes and random walk dynamics \cite{5589113, 2013ISPM...30d..51S}, which serves as the foundation for our analysis.

Random walks are ubiquitous in nature. They arise whenever small, stochastic perturbations accumulate over time \cite{Chandrasekhar:1943ws}.\footnote{
The title of this paper is motivated by the timeless work \cite{Chandrasekhar:1943ws}.
} 
Brownian motion or the erratic motion of a particle immersed in a fluid is a fundamental phenomenon that underlies stochastic processes \cite{Uhlenbeck:1930zz, 1944BSTJ...23..282R, 1945RvMP...17..323W, 1947AmMM...54..369K}. Observations show that the time-average of the squared displacement of the particle grows linearly with observation time -- a result explained by Einstein \cite{1905AnP...322..549E} and Langevin \cite{Langevin1908, 10.1119/1.18725} that was among the first convincing pieces of evidence for the existence of molecules and the molecular composition of matter. This linear growth reflects the intrinsically nonstationary nature of diffusive processes. Pulsar timing observables share this feature: residuals accumulate stochastic effects from spin irregularities, environmental forces, or GWs, leading to diffusive behavior. From this viewpoint, Gaussian processes employed in PTA analyses can be understood as limits of underlying random walk dynamics.

Langevin's approach, which treats a system's dynamics as a deterministic evolution perturbed by rapidly fluctuating stochastic forces, is ideally suited to the problems addressed in this work. In this framework the dynamics are described by Langevin equations (stochastic differential equations, SDEs) in which a deterministic drift term is complemented by a random noise term whose correlation time is much shorter than the characteristic time scales of the system or of the observation \cite{Chandrasekhar:1943ws, 2010kvsp.book.....K}. The method becomes indispensable whenever a purely deterministic description breaks down, such as in the Brownian motion of a particle in a fluid undergoing frequent molecular collisions, the timing residuals produced by a GWB that is the superposition of unresolved sources, and the spin evolution of a pulsar that experiences stochastic internal torques.
A notable feature of this approach is that Gaussianity of the solutions is not assumed but derived. The formal solution of any linear SDE is a stochastic integral of white noise, or a sum of a large number of independent random increments, and Gaussianity follows through the central limit theorem \cite{Chandrasekhar:1943ws, Uhlenbeck:1930zz}. The statistical properties of the process, its means, covariances, and probability density functions, are consequences of forces that enter the equation of motion, not of a prescribed kernel.

Langevin equations have also started to play a key position in pulsar timing analysis through their connection with state space methods. In particular, the prediction step of the Kalman filter \cite{Kalman:434680} is equivalent to solving a Langevin equation that governs the time evolution of the system. This has been put to good use in recent works on pulsar timing glitch \cite{Melatos:2020rlu, Dunn:2023uqo} and noise analysis \cite{Meyers:2021myb, Meyers:2021slh, ONeill:2024uiw, Dong:2025nho}, and in PTA searches for GWs and a GWB \cite{Kimpson:2024fel, Kimpson:2024kgq, Kimpson:2025lju, 2025arXiv251011077K}. The appeal of this approach lies in its computational efficiency: state space algorithms scale linearly with the number of observations, making them more efficient than traditional Gaussian process methods, which scale cubically \cite{5589113, 2013ISPM...30d..51S}. More broadly, the analytical structure of SDE solutions underpins both Kalman-filter state space methods and scalable Gaussian process algorithms such as \texttt{celerite} \cite{2014ApJ...788...33K, 2017AJ....154..220F, 2018PhRvE..98a2136S}, motivating a careful analytical treatment of the SDEs relevant to pulsar timing.

This work takes a step back from numerical implementations to ask what the underlying SDEs reveal about pulsar timing signals,  specifically, by deriving their analytical solutions: means, covariances, and probability density functions. By analytical solutions we mean time-domain expressions; frequency-domain expressions follow more directly by Fourier transforming SDEs and are generally better known.
The body of state space and scalable Gaussian process implementations in pulsar timing is growing rapidly, but the analytical solutions and physical interpretation of the underlying SDEs have not been systematically developed. This work fills that gap. It proceeds more deliberately than the current pace of algorithmic development,\footnote{This will be pursued elsewhere.} but with rigour, providing the explicit solutions and physical understanding that state space methods implicitly rely upon.
This work is structured as follows: we begin by clarifying key concepts (averages, stationarity and ergodicity) that are relevant to interpreting our results (Section \ref{sec:averages}) and examining pulsar timing observables in response to GWs, a stochastic GWB, and intrinsic timing noise (Section~\ref{sec:pulsar_timing_response_to_gwb}). We then introduce Langevin equations as a basis for modeling stochastic processes (Section~\ref{subsec:a_brief_review_of_langevin_equations}), followed by the derivation of analytical solutions to specific SDEs relevant to pulsar timing, including the Ornstein-Uhlenbeck and Mat\'ern processes, and a neutron star two-component model with spin wandering (Sections~\ref{subsec:free_brownian_motion}-\ref{subsec:intrinsic_pulsar_noise}). These analytical solutions are utilized to gain physical insights into the modeling of timing noise and GW signals (Section~\ref{sec:discussion}), culminating in a summary of our findings, future research directions, and open questions (Section~\ref{sec:conclusions}).

The appendices give supporting material. Appendix~\ref{sec:a_useful_lemma_on_random_walks} derives a useful lemma on the probability distribution of random variables~\cite{Chandrasekhar:1943ws}. Appendix~\ref{sec:cov_phase_2cm} contains integral identities necessary for calculating the covariance of the pulsar phase in the two-component model with spin wandering. Additionally, Appendix~\ref{sec:nondimensionalization_of_the_two_component_model} nondimensionalizes the equations of motion to facilitate numerical integration. Appendix~\ref{sec:GPs_and_SDEs} elaborates on the equivalence between Gaussian processes and SDEs. Appendix \ref{sec:generating_matern} illustrates how stationary Mat\'ern processes are generated starting with the OU process.

Throughout this paper, we adopt the metric signature $(-,+,+,+)$ and geometrized units $c = G_{\rm N} = 1$ where $G_{\rm N}$ is Newton's gravitational constant and $c$ is the speed of light in vacuum. We also adopt a bilinear form to express likelihoods, which is equivalent to the full quadratic form commonly used. See the section `Bilinear and quadratic forms' in Appendix~\ref{sec:a_useful_lemma_on_random_walks}.

\section{Time averages, stationarity and ergodicity}
\label{sec:averages}

We briefly review key concepts of ensemble and time averages, stationarity, and ergodicity that will be important to the physical interpretation of the results of this paper. Consider a random process, $z(t)$, understood as a collection of independent realizations (an ensemble) generated by the same underlying probability law.

We begin by defining what is a system and an ensemble. A system is a collection of physical entities that we are analyzing, while an ensemble is a large collection of systems that are identical in their macroscopic properties but differ in their microscopic configurations. 
In the context of pulsar timing, a realization is a single observed time series produced by one instance of the underlying stochastic process. For a GWB, each realization corresponds to a universe populated by a specific draw of source parameters from the underlying astrophysical distribution, and the ensemble is the collection of all such universes. For intrinsic timing noise, each realization is the noise history of one neutron star, and the ensemble is the collection of all neutron stars sharing the same macroscopic properties. In realistic PTA data both signals coexist: a single pulsar time series is a joint realization, one draw from the GWB ensemble and one from its own intrinsic noise ensemble. Crucially, the GWB realization is shared across all pulsars in the same universe, whereas intrinsic noise is independent from pulsar to pulsar. This is precisely why spatial correlations between pulsars, embodied in the Hellings-Downs curve \cite{Hellings:1983fr} and its variance \cite{Allen:2022dzg}, are a definitive signature of the GWB. Since we observe only one realization of each process, ergodicity is the key assumption that allows time averages to stand in for ensemble averages.

The `ensemble average' at time $t$ is
\begin{equation}
    \langle z(t) \rangle = \int dz \, p(z, t) \, z \, ,
\end{equation}
where $p(z,t)$ is the probability density of $z$ at time $t$. The two-point covariance function is
\begin{equation}
    K(t, t') \equiv \langle z(t)\, z(t') \rangle - \langle z(t) \rangle \langle z(t') \rangle .
\end{equation}
The `time average' over an observation window $t \in [t_0, t_0 + T]$ for a single realization is
\begin{equation}
    \bar{z}_T = \frac{1}{T} \int_{t_0}^{t_0 + T} dt'\, z(t') \,  .
\end{equation}

A process is `stationary' if its mean is constant, $\langle z(t) \rangle = \mu$, and its covariance depends only on the lag $|t - t'|$,
\begin{equation}
    K(t, t') = K(|t - t'|) \,,
\end{equation}
so that the statistical properties are invariant under time translation.\footnote{These conditions on the time-invariance of the first two moments are for `weak' or `wide-sense' stationarity. When the full probability distribution is time-invariant, the process is said to be `strictly' or `strongly' stationary.} For a stationary process, the Wiener--Khinchin theorem guarantees that the power spectral density (PSD) and the covariance function form a Fourier transform pair.
This is the operational basis for spectral modeling of timing noise and GWB signals in PTA analyses.

A process is `ergodic' if time averages converge to ensemble averages in the limit of long observation times. Ergodicity is an additional property that some stationary processes possess; stationarity alone does not imply it.

A process that is not stationary is called `nonstationary'. The hallmark of nonstationarity in diffusive processes is a variance that grows with time.
For linear growth, this is precisely the mean square displacement trend observed in the Brownian motion of a particle suspended in a fluid. Several of the processes analyzed in this work, including the Ornstein--Uhlenbeck spin frequency and the two-component residual model, will be shown to be nonstationary in this sense.

\section{Pulsar timing response to a GWB}
\label{sec:pulsar_timing_response_to_gwb}

In this section, we review the timing response of a pulsar to a GW, a GWB, and pulsar intrinsic noise \cite{Maggiore:2018sht}.
This section is kept to make the paper self-contained, and experts in PTA theory may wish to advance to Section \ref{sec:langevin_eqs_for_pulsar_timing}.

\subsection{Pulsar redshift}
\label{subsec:pulsar_redshift}

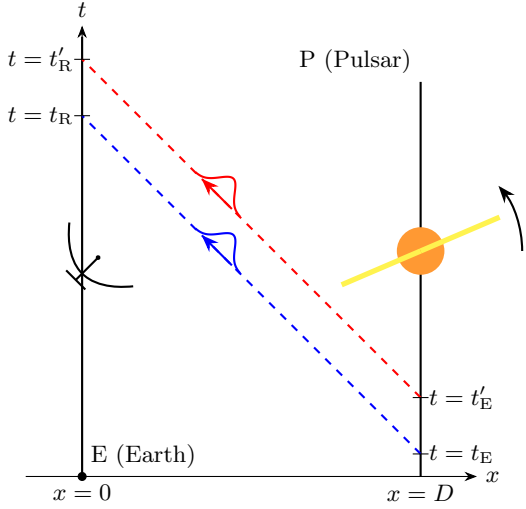
\begin{figure}[t]
\centering
\begin{tikzpicture}[scale=1.5,>=Stealth]

\draw[->] (-0.5,0) -- (3.5,0) node[right] {$x$};
\draw[->] (0,0) -- (0,4) node[above] {$t$};

\draw[thick] (0,0) -- (0,3.9);
\draw[thick] (3,0) -- (3,3.5) node[above left] {P (Pulsar)};

\filldraw (0,0) circle (1pt) node[above right] {E (Earth)};
\node[below] at (0,0) {$x=0$};
\node[below] at (3,0) {$x=D$};

\draw[thick,blue,dashed] (3,0.2) -- (0,3.2);
\draw[thick,red,dashed]  (3,0.7) -- (0,3.7);

\draw[thin] (-0.07,3.2) -- (0.07,3.2); \node[left] at (0,3.2) {$t = t_{\rm R}$};
\draw[thin] (-0.07,3.7) -- (0.07,3.7); \node[left] at (0,3.7) {$t = t_{\rm R}'$};
\draw[thin] (2.93,0.2)  -- (3.07,0.2); \node[right] at (3,0.2) {$t = t_{\rm E}$};
\draw[thin] (2.93,0.7)  -- (3.07,0.7); \node[right] at (3,0.7) {$t = t_{\rm E}'$};

\begin{scope}[shift={(1.2,2.0)}, rotate=135]
    \draw[thick,blue,samples=50,domain=-0.28:0.28]
        plot (\x, {-0.20*exp(-\x*\x/0.018)});
    \draw[->,blue,thick] (-0.18,0) -- (0.20,0);
\end{scope}

\begin{scope}[shift={(1.2,2.5)}, rotate=135]
    \draw[thick,red,samples=50,domain=-0.28:0.28]
        plot (\x, {-0.20*exp(-\x*\x/0.018)});
    \draw[->,red,thick] (-0.18,0) -- (0.20,0);
\end{scope}

\begin{scope}[shift={(3,2.0)}]
    \fill[orange!80] (0,0) circle (6pt);
    \draw[yellow!80, line width=2pt] (0,0) -- +(0.7,0.3);
    \draw[yellow!80, line width=2pt] (0,0) -- +(-0.7,-0.3);
    \draw[->, thick] (0.9,0) arc[start angle=0, end angle=40, radius=0.9];
\end{scope}

\begin{scope}[shift={(0,1.8)}, scale=0.4, rotate=-45]
    \draw[thick] (0,0) parabola (1,0.6);
    \draw[thick] (0,0) parabola (-1,0.6);
    \draw[thick] (0,0) -- (0,0.5);
    \fill (0,0.5) circle (1.5pt);
    \draw[thick] (0,-0.2) -- (0,0);
    \draw[thick] (-0.3,-0.2) -- (0.3,-0.2);
\end{scope}

\end{tikzpicture}
\caption{Spacetime diagram showing consecutive electromagnetic pulses that were emitted by a pulsar (P)
and are headed toward Earth (E) at distance $D$. The pulses are emitted at times $t=t_{\rm E}$
and $t=t_{\rm E}'$ and arrive on Earth at reception times $t=t_{\rm R}$ and
$t=t_{\rm R}'$, respectively.}
\label{fig:spacetime_earth_pulsar}
\end{figure}

Consider two consecutive pulses ($f = {\cal O}\left( 10 \right)$ MHz) emitted by a millisecond pulsar at a distance $D = {\cal O}\left( 1 \right)$ kpc with respect to Earth. Due to the lighthouse effect, the interval between the pulses will be $T_{\rm em} = {\cal O}\left( 10^{-2} \right)$ s in the frame of the pulsar. The interval between consecutive emissions $T_{\rm em}$ will be equal to the interval between consecutive receptions of the signal; since the Earth and the pulsar are at rest relative to each other. However, a GW passing between the line-of-sight of the Earth and the pulsar will interact with the pulses and change the interval between consecutive receptions. Because the frequency of GWs are $f_{\rm gw} = {\cal O}\left( 1 \right)$ yr$^{-1}$ and the frequency of electromagnetic emission are $f_{\rm em} = T_{\rm em}^{-1} = {\cal O}\left( 10^2 \right)$ Hz, geometric optics (geodesic equation) can be expected to be a suitable leading-order approximation to the pulsar timing response due to the presence of a GWB. Figure \ref{fig:spacetime_earth_pulsar} depicts the spacetime trajectories of two consecutive pulses; e.g., that were emitted by a pulsar and were received on Earth.

To derive the pulsar timing response, we work in the transverse-traceless (TT) gauge;
\begin{equation}
\label{eq:metric_tt_gauge}
    ds^2 = -dt^2 + \left( \delta_{ij} + h_{ij}(t, \Vec{x}) \right) dx^i dx^j \,,
\end{equation}
where $h_{ij}(t, \Vec{x})$ is a GW in the TT gauge.
Then assuming that both the pulsar and the Earth are on the $x$ axis, we can write down
\begin{equation}
    dx = - \dfrac{dt}{ \sqrt{1 + h_{xx} } } \simeq - dt \left( 1 - \dfrac{1}{2} h_{xx} \right) \,,
\end{equation}
as the distance traveled by a pulse in an infinitesimal time $dt$. The minus sign is because the pulse is propagating to the negative $x$ direction (Figure \ref{fig:spacetime_earth_pulsar}). Integrating this from emission to reception gives the distance between the pulsar and the Earth;
\begin{equation}
\begin{split}
    D = & \int_{\rm E}^{\rm R} (-dx) \\
    = & \, t_{\rm R} - t_{\rm E} - \dfrac{1}{2} \int_{t_{\rm E}}^{t_{\rm R}} dt' h_{xx}\left(t', \vec{x}\left(t'\right)\right) \,. 
\end{split}
\end{equation}
Labels E and R denote the emission (at the pulsar) and reception (on Earth) events, respectively.
Generalizing the pulsar direction ($\hat{x} \rightarrow \hat{n}_a$; for pulsar $a$) and by utilizing the unperturbed path of a pulse ($t_{\rm R} = t_{\rm E} + D_a$ and ${\vec x}(t) = \left(t_{\rm E} + D_a - t \right) \hat{n}_a$) in the perturbation term, we obtain
\begin{equation}
    t_{\rm R} = t_{\rm E} + D_a + \dfrac{n_a^i n_a^j}{2} \int_{t_{\rm E}}^{t_{\rm E} + D_a} dt' h_{ij}\left(t',\left(t_{\rm E} + D_a - t' \right) \hat{n}_a\right) \,.
\end{equation}
The same calculation can be pursued for a pulse emitted at an interval $T_{\rm em}$ after $t_{\rm E}$. This pulse will be received at Earth at a time
\begin{equation}
\begin{split}
    t_{\rm R}' =  t_{\rm E} + T_{\rm em} + & D_a + \dfrac{n_a^i n_a^j}{2} \int_{t_{\rm E} + T_{\rm em}}^{t_{\rm E} + T_{\rm em} + D_a} dt' \\
    & h_{ij}\left(t',\left(t_{\rm E} + T_{\rm em} + D_a - t' \right) \hat{n}_a\right) \,.
\end{split}
\end{equation}
Shifting the origin of the time coordinate
inside the integral as $t' \rightarrow t' + T_{\rm em}$, we are able to write this as
\begin{equation}
\begin{split}
    t_{\rm R}' =  t_{\rm E} + T_{\rm em} + & D_a + \dfrac{n_a^i n_a^j}{2} \int_{t_{\rm E}}^{t_{\rm E} + D_a} dt' \\
    & h_{ij}\left(t' + T_{\rm em},\left(t_{\rm E} + D_a - t' \right) \hat{n}_a\right) \,.
\end{split}
\end{equation}
The interval between the reception times is given by
\begin{equation}
\begin{split}
    \delta t_{\rm R} = \, & t_{\rm R}' - t_{\rm R} \\
    \delta t_{\rm R} = \, & T_{\rm em} + \dfrac{n_a^i n_a^j}{2} \int_{t_{\rm E}}^{t_{\rm E} + D_a} dt' \\
    & \bigg( h_{ij}\left(t' + T_{\rm em},\left(t_{\rm E} + D_a - t' \right) \hat{n}_a\right) \\
    & \ \ - h_{ij}\left(t',\left(t_{\rm E} + D_a - t' \right) \hat{n}_a\right) \bigg) \,.
\end{split}
\end{equation}

Now, the period of GWs of relevance in PTAs are much longer compared to the interval between the pulses, the last result can be expressed approximately
\begin{equation}
\begin{split}
    \dfrac{\delta t_{\rm R}}{T_{\rm em} } -1 = \, \dfrac{n_a^i n_a^j}{2} & \int_{t_{\rm E}}^{t_{\rm E} + D_a} dt' \\
    & \frac{\partial}{\partial t'} h_{ij}(t', \vec{x}) |_{\vec{x}=\left(t_{\rm E} + D_a - t' \right) \hat{n}_a} \,.
\end{split}
\end{equation}
If there were no GWs in the line-of-sight between the Earth and the pulsar, $\delta t_{\rm R} = T_{\rm em}$; in other words, the pulses would arrive at the same rates on Earth as they were emitted at the pulsar at a distance $D$.

It is convenient to express the result in terms of the pulsar redshift; which is directly related to the pulsar timing residual. This turns out to be
\begin{equation}
\label{eq:pulsar_redshift_sachs_wolfe}
    z_a(t) =  \dfrac{n_a^i n_a^j}{2} \int_{t_{\rm E}+t}^{t_{\rm E} + t + D_a} dt' \frac{\partial}{\partial t'} h_{ij}(t', \vec{x}) |_{\vec{x}=\left(t_{\rm E} + D_a - t' \right) \hat{n}_a} \,,
\end{equation}
where $t$ denotes the laboratory time. This is the expression often considered in the literature to derive the pulsar response function and the pulsar timing signal due to a GWB \cite{Sachs:1967er, Maggiore:2018sht, Romano:2023zhb}. In practice, the pulsar timing residual, a time series $r_{a}(t)$, is utilized. This is related to the pulsar redshift as
\begin{equation}
\label{eq:redshift_to_residual}
    r_a(t) = r_{a,0} + \int_{t_0}^t dt' \, z_a(t') \,.
\end{equation}
The lower limit sets an arbitrary initial phase to the timing residual, $r_{a,0} \equiv r_a(t_0)$.

For a monochromatic GW, $h_{ij}(t, \vec{x}) = {\cal A}_{ij}(\hat{\Omega}) \cos (2\pi f_{\rm gw} (t - \hat{\Omega}\cdot \vec{x}))$, that is propagating along the $\hat{\Omega}$ direction, the pulsar redshift \eqref{eq:pulsar_redshift_sachs_wolfe} can be written as
\begin{equation}
\label{eq:pulsar_redshift_earth_pulsar_terms}
    z_a(t) \equiv \dfrac{n_a^i n_a^j}{ 2\left( 1 + \hat{\Omega} \cdot \hat{n}_a \right) } \left[ h_{ij}\left(t, \vec{0}\right) - h_{ij}(t - D_a, D_a \hat{n}_a) \right] \,.
\end{equation}
The first `Earth' term is the GW perturbation to the pulse geodesic at Earth. The second term is called the pulsar term, and is the GW perturbation to the pulse at the location of the pulsar.

\subsection{GWB signal}
\label{subsec:gwb_signal}

Consider an isotropic, unpolarized, and Gaussian GWB \cite{Bernardo:2024bdc}; defined as a
superposition
of GWs
\begin{equation}
\label{eq:gwb_sum}
\begin{split}
    h_{ij}(t,\vec{x}) = \sum_{A={+,\times}} & \int_{-\infty}^\infty df \int_{\mathbb{S}^2} d\hat{\Omega} \, \\
    & \tilde{h}^A (f, \hat{\Omega})  \epsilon^A_{ij}(\hat{\Omega}) e^{-2\pi i f (t - \vec{x}\cdot \hat{\Omega})} \,,
\end{split}
\end{equation}
where $\epsilon^{A=+,\times}_{ij}(\hat{\Omega})$ are GW polarization basis tensors and the Fourier amplitudes satisfy the ensemble averages
\begin{equation}
\label{eq:h_ave}
    \langle \tilde{h}^A (f, \hat{\Omega}) \rangle = 0 \,,
\end{equation}
\begin{equation}
\label{eq:hh_ave}
    \langle \tilde{h}^A (f, \hat{\Omega}) \tilde{h}^{A'} (f', \hat{\Omega}') \rangle = 0 \,,
\end{equation}
\begin{equation}
\label{eq:hh_powerspectrum}
    \langle \tilde{h}^A (f, \hat{\Omega}) \tilde{h}^{A'*} (f', \hat{\Omega}') \rangle = \dfrac{H(f)}{2} \delta_{AA'} \delta(f - f') \delta( \hat{\Omega}, \hat{\Omega}' ) \,,
\end{equation}
while the higher-point functions satisfy Isserlis' theorem \cite{Isserlis_Theorem}. The reality condition on $h_{ij}(t,\vec{x})$ imposes $\tilde{h}^{A*} (f, \hat{\Omega}) = \tilde{h}^A (-f, \hat{\Omega})$. The quantity $H(f)$ is the power spectrum of the GWB signal; we also choose that it satisfies $H(f)=H(-f)$. The statistics (\ref{eq:h_ave}-\ref{eq:hh_powerspectrum}) and Isserlis' theorem \cite{Isserlis_Theorem} for the higher point functions and moments is consistent with a GWB that is produced by a very large number of individually irresolvable, weak GW sources \cite{Xue:2024qtx}.

Substituting \eqref{eq:gwb_sum} into \eqref{eq:pulsar_redshift_sachs_wolfe}, and performing the time integration, we obtain
\begin{equation}
\label{eq:pulsar_redshift_gwb_sum_tE}
\begin{split}
z_a(t)=& \int_{-\infty}^\infty df \int_{\mathbb{S}^2} d\hat{\Omega} \sum_{A={+,\times}} \tilde{h}^A (f, \hat{\Omega}) F^A_a(\hat{\Omega}) \\
& \,\,\,\, \times U_a(f, \hat{\Omega}) e^{-2\pi i f\left[ t + t_{\rm E} + D_a \right]} \,,
\end{split}
\end{equation}
where the quantities $F^{+/\times}_a(\hat{\Omega})$ are so-called antenna pattern functions;
\begin{equation}
    F_a^A(\hat{\Omega})=\dfrac{ n_a^i n_a^j \epsilon^A_{ij}(\hat{\Omega}) }{2\left(1 + \hat{n}_a\cdot\hat{\Omega}\right)} \,,
\end{equation}
and the $U_a(f, \hat{\Omega})$'s are given by
\begin{equation}
\label{eq:Ua_def}
    U_a(f, \hat{\Omega}) = 1 - e^{2\pi i f D_a (1 + \hat{n}_a\cdot\hat{\Omega})} \,.
\end{equation}
The antenna pattern functions depend solely upon the relative orientation of the Earth-pulsar system to the direction of a GW. The first term of \eqref{eq:Ua_def} corresponds to the Earth term contributions to the pulsar redshift, the second term to the pulsar term.
Following \eqref{eq:h_ave}, it can be shown that
\begin{equation}
\label{eq:redshift_first_moment}
\langle z_a(t) \rangle = 0 \,.
\end{equation}
The quantity of interest is therefore the pulsar redshift correlation;
\begin{equation}
    \langle z_a(t) z_b(t') \rangle \neq 0 \,.
\end{equation}
For a Gaussian signal, this will be all that there is to know; all higher-point functions and cumulants are determined by the two-point function due to Isserlis' theorem \cite{Isserlis_Theorem}. We proceed to calculating this signal.

To simplify the analysis, we recalibrate the lab time, as $t \rightarrow t + t_{\rm E} + D_a$; this is equivalent to fixing the phase of the waveform so that $t_{\rm E} = -D_a$ and other values of $t_{\rm E}$ corresponds to shifting the initial phase\footnote{When computing the correlation between a pair of pulsars $a$ and $b$, this initial phase information automatically drops off for $D_a=D_b$; attesting to its nonobservability.}. Thus, we will work with the expression for the pulsar redshift;
\begin{equation}
\label{eq:pulsar_redshift_gwb}
\begin{split}
z_a(t)=& \int_{-\infty}^\infty df \int_{\mathbb{S}^2} d\hat{\Omega} \sum_{A={+,\times}} \tilde{h}^A (f, \hat{\Omega}) F^A_a(\hat{\Omega}) \\
& \,\,\,\, \times U_a(f, \hat{\Omega}) e^{-2\pi i f t}  \,.
\end{split}
\end{equation}
The correlation is given by
\begin{align}
\langle z_a(t) z_b(t') \rangle = & \int df d\hat{\Omega} \, df' d\hat{\Omega}' \sum_{AA'} e^{-2\pi i (f t - f' t')}  \nonumber \\
& \times F^A_a(\hat{\Omega}) U_a(f, \hat{\Omega}) F^{A'*}_b(\hat{\Omega}') U_b^*(f', \hat{\Omega}') \nonumber \\
& \times \langle \tilde{h}^A (f, \hat{\Omega}) \tilde{h}^{A'*} (f', \hat{\Omega}') \rangle \phantom{\dfrac{1}{1}} \,.
\label{eq:redshift_correlation_init}
\end{align}
Utilizing \eqref{eq:hh_powerspectrum}, and performing the integrals over $f'$ and $\hat{\Omega}'$ as well as the sum over $A'$, we obtain
\begin{equation}
\label{eq:redshift_correlation_init2}
\begin{split}
\langle z_a(t) z_b(t') \rangle = & \int df d\hat{\Omega} \sum_A e^{-2\pi i f (t - t')} \dfrac{H(f)}{2}  \\
& \times F^A_a(\hat{\Omega}) F^{A*}_b(\hat{\Omega}) U_a(f, \hat{\Omega})  U_b^*(f, \hat{\Omega}) \,.
\end{split}
\end{equation}
Defining the overlap reduction function (ORF);
\begin{equation}
\label{eq:orf_def}
\gamma_{ab}(f) = \sum_{A=+,\times} \int_{\mathbb{S}^2} d\hat{\Omega} \, F^A_a(\hat{\Omega}) F^{A*}_b(\hat{\Omega}) U_a(f, \hat{\Omega})  U_b^*(f, \hat{\Omega}) \,,
\end{equation}
we have
\begin{equation}
\label{eq:redshift_correlation_general}
\begin{split}
\langle z_a(t) z_b(t') \rangle = & \int_{-\infty}^\infty df \, e^{-2\pi i f (t - t')} \dfrac{H(f)}{2} \gamma_{ab}(f) \,.
\end{split}
\end{equation}
In the long arm limit, $D/\lambda_{\rm gw} \gg 1$, the angular integration can be carried out analytically {\it a la} Hellings \& Downs \cite{Hellings:1983fr, 1985MNRAS.217..265M}, leading to
\begin{equation}
    \gamma_{ab}(f) \sim \Gamma_{ab} \,, \,\,\,\, {\rm as} \,\,\,\, D/\lambda_{\rm gw} \gg 1 \,,
\end{equation}
where $\Gamma_{ab}$ is the HD curve; for $a \neq b$,
\begin{equation}
\label{eq:hd_curve}
    \Gamma_{ab} =  \dfrac{1}{2} + \dfrac{3}{4} \left(1-\hat{n}_a \cdot \hat{n}_b\right) \left[ \ln \left( \dfrac{1-\hat{n}_a \cdot \hat{n}_b}{2} \right) - \dfrac{1}{6} \right] \,.
\end{equation}
For PTA applications, pulsars have $D={\cal O}(1)$ kpc and the GWs of SMBHBs are $\lambda_{\rm gw} = {\cal O}(1)$ pc.
The long arm limit is a valid approximation. It is worth noting that in this limit, the frequency dependence of the correlation drops completely.\footnote{Finite pulsar distance effects become relevant at low frequencies $f \sim 0.1 \, {\rm yr^{-1}}$, pulsar distances $D \lesssim 100 \, {\rm pc}$, and pulsar pairs with sub-degree angular separations \cite{Ng:2021waj, Allen:2022dzg, Domenech:2024pow}.} For $a = b$, the autocorrelation is obtained by $\gamma_{aa} \equiv 2 \Gamma_{aa}$; the factor of 2 is due to small scale power \cite{Ng:2021waj, Domenech:2024pow, Hu:2022ujx}. To a good approximation, the pulsar redshift correlation can be written as\footnote{When dealing with timing residuals \eqref{eq:redshift_to_residual}, the corresponding two point correlation can be obtained by replacing the PSD with $H(f)/(2\pi f)^2$.}
\begin{equation}
\label{eq:redshift_correlation_gwb_longarm}
\begin{split}
\langle z_a(t) z_b(t') \rangle = \, & \Gamma_{ab} \int_{0}^\infty df \cos\left({2\pi f (t - t')}\right) H(f) \,.
\end{split}
\end{equation}
For $a=b$ (same pulsar), the result shows that the leading order PTA signal of a GWB can be viewed as a time correlated, stationary, Gaussian random process (via the Wiener-Khinchin theorem) with a PSD $H(f)$ \cite{1985MNRAS.217..265M}. For $a \neq b$ (different pulsars), the signal is spatially correlated with a covariance determined by the HD curve.

\subsection{Time correlated noises}
\label{subsec:correlated_noises}

The GWB signal \eqref{eq:redshift_correlation_gwb_longarm} is not the only correlated process in pulsars \cite{1982MNRAS.199..659M, 1985MNRAS.217..265M, vanHaasteren:2008yh, Coles:2011zs, vanHaasteren:2012hj}. Intrinsic red noise is particularly relevant, and a dominant contribution to timing noise in pulsars \cite{Melatos:2013rca, Haskell:2015jra, Goncharov:2019mei}. Nonetheless, via the Wiener-Khinchin theorem, we can associate the noise contribution to pulsars to have the following leading properties;
\begin{equation}
\label{eq:redshift_mean_noise}
\begin{split}
\langle z_a(t) \rangle = 0 \,,
\end{split}
\end{equation}
\begin{equation}
\label{eq:redshift_correlation_noise}
\begin{split}
\langle z_a(t) z_b(t') \rangle = \, & \delta_{ab} \int_{0}^\infty df \cos\left({2\pi f (t - t')}\right) S_a(f) \,,
\end{split}
\end{equation}
where $S_a(f)$ is a PSD that is determined for each pulsar. The Kronecker delta indicates that noise is uncorrelated across pulsars. Then, the covariance in PTA analysis is taken as a sum of the signal \eqref{eq:redshift_correlation_gwb_longarm} and noise \eqref{eq:redshift_correlation_noise} components.

It is instructive to compare the pulsar redshift covariances \eqref{eq:redshift_correlation_gwb_longarm} and \eqref{eq:redshift_correlation_noise}. The GWB signal is characterized by a single PSD $H(f)$ and the HD correlation $\Gamma_{ab}$ \eqref{eq:hd_curve}. On the other hand, intrinsic pulsar noises are determined by a PSD $S_a(f)$ that is characteristic to each pulsar and uncorrelated spatially. Both GWB signal and pulsar noises are characterized by zero means and covariance matrices that can be translated into each other (mathematically) as $\Gamma_{ab }H(f) \rightleftarrows \delta_{ab} S_a(f)$. Results that would be derived later (Sections \ref{subsec:a_brief_review_of_langevin_equations}-\ref{subsec:brownian_oscillator}) hold for both the GWB signal and pulsar noises, even when the discussion will be focused on one.

We will also show that pulsar timing noise (correlated red noise) physically arises via 
stochastic torques acting on a neutron star (Section~\ref{subsec:intrinsic_pulsar_noise}). 
Such torques generally render the timing residuals nonstationary, and recognizing this is 
crucial for a precise analysis.

\section{Langevin equations for pulsar timing}
\label{sec:langevin_eqs_for_pulsar_timing}

In this section, we briefly review Langevin equations and show explicit examples that derive analytical time-domain solutions (means, covariances and PDFs of pulsar redshifts and timing residuals).

\subsection{A brief review of Langevin equations}
\label{subsec:a_brief_review_of_langevin_equations}

We are concerned with an $n$th order linear time invariant stochastic differential equation (SDE);
\begin{equation}
\label{eq:sde_lte}
    \left( \dfrac{d^n}{dt^n} + a_{n-1} \dfrac{d^{n-1}}{dt^{n-1}} + \cdots + a_1 \dfrac{d}{dt} + a_0 \right) y(t) = w(t) \,,
\end{equation}
where $\{ a_0 , \cdots a_{n-1} \}$ is a set of $n$ constants and $w(t)$ is a white Gaussian noise with a PSD $s$, zero mean and a covariance
\begin{equation}
\label{eq:wn_cov_sde_lte}
    \langle w(t) w(t') \rangle = \dfrac{s}{2} \delta(t - t') \,.
\end{equation}
It is worth noting that a SDE \eqref{eq:sde_lte} can always be written for a given Gaussian process, supporting an equivalence of both descriptions for random phenomena \cite{5589113, 2013ISPM...30d..51S}. We recommend Appendix \ref{sec:GPs_and_SDEs} to readers interested to dwell deeper into the subtleties of this equivalence.

Langevin equations are SDEs that describe how a system responds to deterministic and stochastic forces \cite{Langevin1908, Chandrasekhar:1943ws}. The simplest of these is that of a free Brownian particle; in one dimension, this is described by the equation of motion of a particle of mass $m$, position $x$ and velocity $v = dx/dt$;
\begin{equation}
\label{eq:langevin_brownian}
    m \dfrac{d v}{dt} = - b v + \xi(t) \,,
\end{equation}
where the stochastic driving force $\xi(t)$ has zero mean and a covariance
\begin{equation}
    \langle \xi(t) \xi(t') \rangle = \sigma_{\rm F}^2 \delta( t - t' ) \,.
\end{equation}
The drag force on the right hand side of \eqref{eq:langevin_brownian} is due to dynamical friction between the particle and the fluid or molecules in the fluid; whereas its drag coefficient $b$ can be determined by Stokes' law \cite{Chandrasekhar:1943ws}. The fluctuating part $\xi(t)$ is assumed to vary extremely rapidly compared to variations of macroscopic observables. The solution of \eqref{eq:langevin_brownian} has been shown to describe precisely the dynamics of a microscopic particle that suffers $\sim 10^{21}$ collisions per second (experimental time scale), consistent with Einstein's solution in terms of a diffusion equation \cite{1905AnP...322..549E}. We will return to this equation and its solution in the context of PTAs in the next section (Section \ref{subsec:free_brownian_motion}).

Another Langevin equation that we shall find of importance is that of a Brownian harmonic oscillator \cite{Uhlenbeck:1930zz, Chandrasekhar:1943ws}. The equation of motion is given by
\begin{equation}
\label{eq:langevin_harmonic}
    m \dfrac{d^2 x}{dt^2} = - b \dfrac{dx}{dt} - k x + {{\mathbf \xi}}(t) \,,
\end{equation}
where the system dynamics is determined by dynamical friction, $-bv$, a restoring force (or Hooke's law with a spring constant $k$), $ -kx$, and a stochastic driving term, $\xi(t)$. The Langevin equation is this time explicitly expressed in the particle's position. Notice that without the stochastic driving term, this is simply the equation of motion of a damped harmonic oscillator. The physical motivation of the Brownian harmonic oscillator is that the dynamics is acted upon by a restoring force, in addition to dynamical friction; we shall find that this feature has implications for modelling and fitting data. We will tease out the solution to this in Section \ref{subsec:brownian_oscillator} in a PTA context.

For applications of Langevin equations outside traditional Brownian movement and astrophysics, we refer readers to \cite{2008PhRvE..77b2101P, Harko:2012yt, Leung:2013iza, 2022Chaos..32l3138M,2007ITED...54.1185J, 2010JEE....61..252V, 2008PhRvL.100g0601B, 2008PhRvE..78c1112B, 2000PhyA..287..482L, 2002PhRvE..65c7106W, 2002PhRvE..66d6118P}.

\subsection{Free Brownian motion}
\label{subsec:free_brownian_motion}

Consider the SDE corresponding to an Ornstein-Uhlenbeck (OU) process \cite{Uhlenbeck:1930zz, Chandrasekhar:1943ws};
\begin{equation}
\label{eq:sde_exp}
    \left( \dfrac{d}{dt} + \dfrac{1}{l} \right) z_a(t) = w_a(t) \,
\end{equation}
where $w_a(t)$ is a white noise process with a mean
\begin{equation}
\label{eq:wnave_exp}
\langle w_a(t) \rangle = 0
\end{equation}
and a covariance
\begin{equation}
\label{eq:wncov_exp}
    \langle w_a(t) w_b(t') \rangle = \Gamma_{ab} \dfrac{ 2\sigma^2}{l} \delta( t - t' ) \,.
\end{equation}
Later the specific functional form used for the white noise covariance function will become transparent when we show that this SDE describes a Gaussian process with an exponential kernel. This form can be reached with \eqref{eq:langevin_brownian} by the identification $z_a(t) = v(t)$, $l = m/b$ and $\Gamma_{ab} \sigma^2 = \sigma_F^2/(2 m b)$. The quantity $l$ takes the meaning of a correlation time scale. In the PTA context, we consider $z_a(t)$ to be the pulsar redshift and $\Gamma_{ab}$ the HD curve.
We will show that the solution of (\ref{eq:sde_exp}-\ref{eq:wncov_exp}) gives the PTA GW signal \eqref{eq:redshift_correlation_gwb_longarm}.

The formal solution of \eqref{eq:sde_exp} can be written. Multiplying both sides of \eqref{eq:sde_exp} by $e^{t/l}$, we obtain
\begin{equation}
    \dfrac{d}{dt} \left( e^{t/l} z_a(t) \right) = w_a(t) e^{t/l} \,.
\end{equation}
Integrating both sides gives
\begin{equation}
\label{eq:formal_solution_exp}
    z_a(t) - z_{a,0} e^{-(t-t_0)/l} = \int_{t_0}^t dt_1 \, w_a(t_1) e^{-(t-t_1)/l} \,,
\end{equation}
where $z_{a,0} = z_a(t_0)$ is constant in time. If this were an ODE, then this would be the end of it. However, $w(t)$ is a random variable whose properties are only determined only on average. Accordingly, the solution of a SDE requires a specification of the $n$-point function $\langle z_1(t_1) z_2(t_2) \cdots z_n(t_n) \rangle$ or the joint PDF of the pulsar redshifts at $n$ different times for $n$ different pulsars. This procedure will be carried out explicitly for $\langle z_a (t) \rangle$ and $\langle z_a (t)z_b (t') \rangle$.

Because $\langle w_a (t) \rangle = 0$, we see that
\begin{equation}
    \langle z_a(t) \rangle = z_{a,0} e^{-(t-t_0)/l} \,.
\end{equation}
The covariance can be derived by first writing
\begin{equation}
\begin{split}
& \langle z_a(t) z_b(t') \rangle \phantom{\dfrac{1}{1}} \\
& \,\, = z_{a,0} z_{b,0} e^{-(t+t'-2t_0)/l} \phantom{\dfrac{1}{1}} \\
& \,\,\,\,\,\,\,\,\,\, + \int_{t_0}^t dt_1 \int_{t_0}^{t'} dt_2 \, e^{-(t+t'-t_1-t_2)/l} \langle w_a(t_1) w_b(t_2) \rangle \,. 
\end{split}
\end{equation}
Substituting the noise covariance \eqref{eq:wncov_exp} and evaluating the integral gives
\begin{equation}
\begin{split}
     \langle z_a(t) z_b(t') \rangle = \, & \Gamma_{ab} \sigma^2 e^{-| t'-t |/l} \\
     & + \left( z_{a,0}z_{b,0} - \Gamma_{ab} \sigma^2 \right) e^{-(t'  + t - 2 t_0)/l} \,.
\end{split}
\end{equation}
The solution is made of a stationary piece $\sim e^{-|t'-t|/l}$ and a nonstationary piece $\sim e^{-(t + t')/l}$ that gets damped exponentially.
At large times compared to the initial time, the nonstationary contribution becomes increasingly subdominant compared to the stationary part; in particular, in the limit $t', t \gg t_0$, we have
\begin{eqnarray}
    \langle z_a(t) \rangle & \rightarrow & 0 \,, \\
    \langle z_a(t) z_b(t') \rangle & \sim & \Gamma_{ab} \sigma^2 e^{-|t'-t|/l} \,.
\end{eqnarray}
This shows that in the large time limit, the OU process corresponds to a stationary Gaussian process with an exponential kernel, i.e., $z_a(t) \sim {\cal N}\left( 0, \Sigma_{ab} \right)$ where $\Sigma_{ab} = \Gamma_{ab} \sigma^2 e^{-|t'-t|/l}$. To fully establish this equivalence, it can be shown that the substitution of the PSD
\begin{equation}
\label{eq:psd_exp_kernel} H(f) = \dfrac{ 4 \sigma^2 l }{ 1 + \left( 2 \pi f l \right)^2 } 
\end{equation}
into \eqref{eq:redshift_correlation_gwb_longarm} gives exactly
\begin{equation}
\label{eq:redshift_correlation_brownian}
\langle z_a(t) z_b(t') \rangle = \Gamma_{ab} \sigma^2 e^{-|t'-t|/l} \,.
\end{equation}
The parameters $\sigma$ and $l$ could now be associated with GWB parameters, i.e., $\sigma \sim A_{\rm gw}^\star$ and $l^{-1} \sim f^\star$ where $f^\star$ is a frequency scale such that for $f \ll f^\star$ the PSD is flat ($H(f)\sim 4 (A_{\rm gw}^\star)^2 / f^\star$) while for $f \gg f^\star$ the PSD varies as a power law $H(f) \sim \left((A_{\rm gw}^\star)^2 / (\pi^2 f^\star)\right) ({f/f^\star})^{-2}$.

We can also obtain the PDF of the pulsar redshift (and even the joint PDF of the redshift and timing residual) as a function of the observation time $t$. To do so, we return to the formal solution of the OU process \eqref{eq:formal_solution_exp}. The right hand side of this expression can be interpreted as a sum of many independent, but nonidentically distributed random steps. By virtue of the central limit theorem, the probability distribution of the resultant sum will be Gaussian. The equality of \eqref{eq:formal_solution_exp} entails that the statistical properties of the resultant is shared by the left hand side. To see this explicitly, we write down the integral as
\begin{equation}
\label{eq:wa_integral_to_sum}
    \int_{t_0}^t dt_1 \, w_a(t_1) e^{-(t-t_1)/l} \sim \sum_{j=1}^{(t-t_0)/\Delta t} {Z}_{a,j}(t) \,
\end{equation}
where $Z_{a,j}(t)$ is given by
\begin{equation}
    Z_{a,j}(t) = \int_{t_0 + (j-1)\Delta t}^{t_0 + j\Delta t} dt_j \, w_a(t_j) e^{-(t - t_j)/l} \,.
\end{equation}
In the limit $\Delta t \rightarrow 0$, the sum approaches the Riemann sum expression of the integral and the left and right hand sides of \eqref{eq:wa_integral_to_sum} become exactly equal. With $\Delta t / (t - t_0) \ll 1$, the problem of determining the statistical properties of $z_a(t) - z_{a,0} e^{-(t-t_0)/l}$ reduces to the problem of a one-dimensional random walk with independent and nonidentically distributed steps $Z_{a,j}(t)$. The independence of $Z_{a,j}(t)$'s follows through the statistics of $w_a(t)$. With the bins sufficiently small, we approximate $Z_{a,j}(t)$ as
\begin{equation}
    Z_{a,j}(t) \sim e^{-(t - \overline{t}_j)/l} \int_{t_0 + (j-1)\Delta t}^{t_0 + j\Delta t} dt_j \, w_a(t_j)  \,,
\end{equation}
where in the exponential we have considered the midpoints $t_j\sim \overline{t}_j= t_0 + \Delta t (j-1/2)$ of each bin.
The assumption that the noise $w_a(t)$ varies extremely rapidly compared to any other time scales justifies the last approximation. Then, clearly,
\begin{eqnarray}
    \langle Z_{a,j}(t) \rangle &=& 0 \,, \\
    \langle Z_{a,j}(t)^2 \rangle &=& \dfrac{2\sigma^2 \Delta t}{l} e^{-2(t - \overline{t}_j)/l} \,.
\end{eqnarray}
Because the steps are independent, the PDF of the resultant $\sum_{j=1}^{(t-t_0)/\Delta t} {Z}_{a,j}(t)$ will have a zero mean and a variance that is the sum of all the independent steps, $\sum_{j=1}^{(t-t_0)/\Delta t} \langle Z_{a,j}(t)^2 \rangle$. The expression for the variance can be calculated as
\begin{equation}
\begin{split}
\sum_{j=1}^{(t-t_0)/\Delta t} \langle Z_{a,j}(t)^2 \rangle & = \sum_{j=1}^{(t-t_0)/\Delta t} \dfrac{2\sigma^2 \Delta t}{l} e^{-2(t - \overline{t}_j)/l}   \\
& \sim \dfrac{2\sigma^2 }{l} \int_{t_0}^t dt_1 \, e^{-2(t - t_1)/l} \,.
\end{split}    
\end{equation}
This gives 
\begin{equation}
    \sum_{j=1}^{(t-t_0)/\Delta t} \langle Z_{a,j}(t)^2 \rangle \sim \sigma^2 \left( 1 - e^{-2(t - t_0)/l} \right) \,.
\end{equation}
The approximations become exact in the limit $\Delta t \rightarrow 0$. Finally, the PDF of the pulsar redshift can be shown to be
\begin{equation}
\begin{split}
    P(z_a, t | z_{a,0}, t_0) \equiv & \dfrac{1}{\sqrt{2\pi \sigma^2 \left( 1 - e^{-2(t - t_0)/l} \right)}} \\
    & \exp\left( - \dfrac{ \left( z_a - z_{a,0} e^{-(t-t_0)/l} \right)^2 }{2 \sigma^2 \left( 1 - e^{-2(t - t_0)/l} \right) } \right) \,.
\end{split}
\end{equation}
At large times compared to the initial time, the explicit time dependence of the PDF completely drops out, leaving $P(z_a, t | z_{a,0}, t_0) \sim e^{-z_a^2/(2\sigma^2)}/\sqrt{2\pi \sigma^2}$. Very near the initial time, the PDF reduces to $P(z_a, t | z_{a,0}, t_0) \sim \delta\left( z_{a} - z_{a,0} \right)$.

We emphasize that in deriving the PDF, we have not assumed that the PDF is Gaussian. Rather, this result followed due to the displacement $z_a(t) - \langle z_a(t) \rangle$ (given by \eqref{eq:wa_integral_to_sum}) treated as a large number of independent random steps $Z_{a,j}(t)$. Then, the central limit theorem gives that the PDF of the sum is Gaussian.

Following similar lines of arguments, we can obtain the joint PDF of the pulsar redshifts for a pair of pulsars $a$ and $b$. The relevant quantity to be calculated to reach this goal is the covariance between independent steps at unequal times. This can be shown to be
\begin{equation}
    \langle Z_{a,j}(t) Z_{b,k}(t') \rangle = \Gamma_{ab} \delta_{jk} \dfrac{2\sigma^2 \Delta t}{l} e^{-(t + t' - \overline{t}_j - \overline{t}_k)/l} \,.
\end{equation}
Then, the covariance of the resultant will be $\sum_{jk} \langle Z_{a,j}(t) Z_{b,k}(t') \rangle$. The sum can be evaluated analytically in the limit $\Delta t \rightarrow 0$ by converting the resulting Riemann sum into its integral form and performing the integration. The resultant covariance is given by
\begin{equation}
\begin{split}
    & \sum_{jk} \langle Z_{a,j}(t) Z_{b,k}(t') \rangle \\ 
    & \,\,\,\, = \Gamma_{ab} \sigma^2 \left(  e^{-|t'-t|/l} - e^{-(t' + t - 2 t_0)/l}  \right) \,.
\end{split}
\end{equation}
The PDF of the redshifts of a pair of pulsars is given by
\begin{equation}
\begin{split}
    & \ln P(z_a, t, z_b, t'| z_{a,0}, z_{b,0}, t_0) \phantom{\dfrac{1}{1}} \\
    & = -\dfrac{1}{2} \ln | 2\pi C^Z_{ab}(t,t',t_0) | \\
    & \phantom{iii} - \dfrac{1}{2} Z[z_a, t, z_{a,0}, t_0] C^Z_{ab}(t,t',t_0)^{-1} Z[z_b, t', z_{b,0}, t_0] \,,
\end{split}
\end{equation}
where the functional $Z$ is given by
\begin{equation}
    Z[z_a, t, z_{a,0}, t_0] = z_a - z_{a,0} e^{-(t-t_0)/l}
\end{equation}
and the covariance is
\begin{equation}
    C^Z_{ab}(t,t',t_0) = \Gamma_{ab} \sigma^2 \left(  e^{-|t'-t|/l} - e^{-(t' + t - 2 t_0)/l}  \right) \,.
\end{equation}
The case $a=b$ and $t=t'$ reduces to the expression for $P(z_a, t | z_{a,0}, t_0)$. For large times compared to the initial time, we obtain
\begin{equation}
\begin{split}
    & \ln P(z_a, t, z_b, t'| z_{a,0}, z_{b,0}, t_0) \\
    & = -\dfrac{1}{2} \ln | 2\pi \Gamma_{ab} \sigma^2 e^{-|t'-t|/l} | - \dfrac{z_a \Gamma_{ab}^{-1}z_b}{2} \dfrac{ e^{|t'-t|/l} }{\sigma^2}  \,.
\end{split}
\end{equation}
This is the Gaussian likelihood of pulsar redshifts $z_a$ at time $t$ and $z_b$ at time $t'$, determined by a time-domain covariance function $C^Z_{ab}(t,t',t_0)\sim \Gamma_{ab} \sigma^2 e^{-|t'-t|/l}$ with a corresponding PSD \eqref{eq:psd_exp_kernel}. The result also confirms that the OU process is a stationary Gaussian process.

The PDF for pulsar timing residuals and the joint PDF of the timing residuals and the redshifts can also be obtained by following the same reasoning that teased out the physics of Brownian motion \cite{Chandrasekhar:1943ws, 2010kvsp.book.....K}. Substituting the formal solution \eqref{eq:formal_solution_exp} into \eqref{eq:redshift_to_residual}, we obtain\footnote{The double integral has been converted into a single integral plus an extra term by an integration by parts. Note that
\begin{equation}
\nonumber
\begin{split}
& \dfrac{d}{dt_2} \left[ -l e^{-t_2/l} \int_{t_0}^{t_2} dt_1\, w_a(t_1) e^{t_1/l} \right] \\
& \,\,\,\, = e^{-t_2/l} \int_{t_0}^{t_2} dt_1\, w_a(t_1) e^{t_1/l} - l w_a(t_2) \,.
\end{split}
\end{equation}
Changing the order of integration is also another way to get to the same result.
}
\begin{equation}
\label{eq:formal_solution_exp_residual}
\begin{split}
    r_a(t) - r_{a,0} & - l z_{a,0} \left( 1 - e^{-(t-t_0)/l} \right) \\
    & = l \int_{t_0}^t dt_1 w_{a}(t_1) \left[ 1 - e^{-(t-t_1)/l} \right] \,.
\end{split}
\end{equation}
The statistical properties of the left hand side should match the right hand side. Then, the right hand side can be recognized as a sum of a large number of independent, nonidentically distributed unbiased random walks. Utilizing the Lemma in Appendix \ref{sec:a_useful_lemma_on_random_walks}, we obtain the likelihood
\begin{equation}
\begin{split}
    & \ln P(r_a, t, r_b, t'| r_{a,0}, z_{a,0}, r_{b,0}, z_{b,0}, t_0) \phantom{\dfrac{1}{1}} \\
    & = -\dfrac{1}{2} \ln | 2\pi C^R_{ab}(t,t',t_0) | \\
    & \phantom{iii} - \dfrac{1}{2} R[z_a, t, z_{a,0}, t_0] C^R_{ab}(t,t',t_0)^{-1} R[z_b, t', z_{b,0}, t_0] \,,
\end{split}
\end{equation}
where the functional $R$ is given by
\begin{equation}
    R[r_a, t, r_{a,0}, z_{a,0}, t_0] = r_a - r_{a,0} - l z_{a,0} \left( 1 - e^{-(t-t_0)/l} \right)
\end{equation}
and the covariance is given by
\begin{equation}
\begin{split}
    C^R_{ab}(t,t',t_0) = 
    \, & \Gamma_{ab} \sigma^2 l^2 \bigg[ -e^{-|t'-t|/l} - e^{-(t'+t-2t_0)/l} \\
    & + 2 e^{-(t' - t_0)/l} + 2 e^{-(t - t_0)/l} \\
    & -2 \left( 1 - \dfrac{ {\rm min}(t,t')-t_0}{l} \right)  \bigg] \,.
\end{split}
\end{equation}
In the limit of large times compared to the initial time, $t', t \gg t_0$, this becomes\footnote{
The function ${\rm min}(t,t')$ can be recognized as a sum of stationary and nonstationary parts: ${\rm min}(t, t') = (t + t' - |t - t'|)/2$.
}
\begin{equation}
\label{eq:timing_residual_cov_freebrownian_steady_state}
\begin{split}
    C^R_{ab}(t,t',t_0) \sim 
    \, & \Gamma_{ab} \sigma^2 l^2 \bigg[ -e^{-|t'-t|/l}  \\
    & -2 \left( 1 - \dfrac{{\rm min}(t,t')-t_0}{l} \right)  \bigg] \,.
\end{split}
\end{equation}
The covariance of pulsar timing residuals increases linearly in time; reminiscent to the mean square displacement of a Brownian particle. The reader may also consult \cite{1996PhRvE..54.2084G} or in an astrophysical context Appendix B of \cite{Antonelli:2022gqw} for related results and discussion on the OU process and its integral.

With this, we have shown that the OU process (Langevin equation for a free Brownian particle (\ref{eq:sde_exp}-\ref{eq:wncov_exp})) gives a consistent description of a PTA signal in the redshifts as a Gaussian process model with an exponential kernel.
Depending on the time scale $l$, the PSD in the pulsar redshift correlation can be effectively flat or a power law $f^{-2}$. A limitation of the OU process is its shallow PSD \eqref{eq:psd_exp_kernel}. Because of this, the traditional interpretation of a GWB due to circular SMBHBs $H(f) \sim f^{-7/3}$ cannot be accommodated within the model \cite{Bernardo:2024bdc, Cai:2020ovp}; see Appendix A of \cite{Kimpson:2025lju}. This issue is relevant to fitting data. Moreover, the signal produced in the timing residuals is nonstationary \cite{1996PhRvE..54.2084G, Antonelli:2022gqw}; due to random walk dynamics, the mean square displacement of the timing data increases linearly. This motivates us to look beyond the OU process for a more flexible basis of the GWB signal (pulsar redshifts and timing residuals) in a PTA.

\subsection{Brownian harmonic oscillator}
\label{subsec:brownian_oscillator}

Consider the SDE \cite{Uhlenbeck:1930zz, Chandrasekhar:1943ws}
\begin{equation}
\label{eq:sde_m32}
    \left( \dfrac{d^2}{dt^2} + \dfrac{2\sqrt{3}}{l} \dfrac{d}{dt} + \left( \dfrac{\sqrt{3}}{l} \right)^2 \right) z_a(t) = w_a(t) \,
\end{equation}
where the noise is characterized by
\begin{equation}
\label{eq:wnave_m32}
\langle w_a(t) \rangle = 0
\end{equation}
and
\begin{equation}
\label{eq:wncov_m32}
    \langle w_a(t) w_b(t') \rangle = \Gamma_{ab} \left( \dfrac{ 12 \sqrt{3 }\sigma^2 }{l^3} \right) \delta( t - t' ) \,.
\end{equation}
This can be reached with the Langevin equation of a Brownian harmonic oscillator \eqref{eq:langevin_harmonic} with the identification, $z_a(t) = x(t)$, $b/m = 2\sqrt{3}/l$, $k/m=(\sqrt{3}/l)^2$, and $\sigma_F^2/m^2 = 12 \sqrt{3} \sigma^2/l^3$. The restoring force $-3 z_a(t)/l^2$ or factor of $\sqrt{3}$ in \eqref{eq:sde_m32} is introduced for convenience to match the usual analytic form of the corresponding Mat\'ern process.

The solution can be written formally.
First, the homogeneous part of the SDE \eqref{eq:sde_m32} has two independent (overdamped oscillator) solutions: $z_a(t) \sim e^{-\sqrt{3}t/l}$ and $z_a(t) \sim  t e^{-\sqrt{3}t/l}$. For simplicity, we consider the initial conditions $z_a(t_0) = 0$ and $z_a'(t_0) = 0$ with $t_0 = 0$. The solution becomes
\begin{equation}
    z_a(t) = \int_0^t dt_1 \, e^{-\sqrt{3} (t - t_1)/l} (t - t_1) w_a(t_1) \,.
\end{equation}
This shows that $\langle z_a(t) \rangle = 0$.
The covariance becomes
\begin{equation}
\begin{split}
    \langle z_a(t) z_b(t') \rangle = \, & e^{-\sqrt{3} (t+t')/l} \int_0^t dt_1 \int_0^{t'} dt_2 \, e^{\sqrt{3}(t_1+t_2)/l} \, \\
    & \times (t - t_1)(t'-t_2) \langle w_a(t_1) w_b(t_2) \rangle \,.
\end{split}
\end{equation}
This simplifies to
\begin{equation}
\begin{split}
     & \langle z_a(t) z_b(t') \rangle \phantom{\dfrac{1}{1}} \\
     & \,\, = \Gamma_{ab} \sigma^2 \bigg[ \left( 1 + \sqrt{3} \dfrac{|t'-t|}{l} \right) e^{-\sqrt{3} |t'-t|/l} \\
     & \phantom{gggggggg} - \left( 1 + 6 \dfrac{t t'}{l^2} + \sqrt{3} \dfrac{t+t'}{l} \right) e^{-\sqrt{3}(t'+t)/l}  \bigg] \,.
\end{split}
\end{equation}
The solution is again a combination of a stationary piece and a nonstationary piece. Nonetheless, at large times compared to the initial time, $t', t \gg t_0$, the nonstationary part becomes subdominant. This leaves us with
\begin{align}
    \langle z_a(t) \rangle & \rightarrow 0 \,, \\
    \langle z_a(t) z_b(t') \rangle & \sim \Gamma_{ab} \sigma^2\left( 1 + \sqrt{3} \dfrac{| t'-t |}{l} \right) e^{-\sqrt{3}|t'-t|/l} \,.
\end{align}
The above shows that the random process $z_a(t)$ -- that is the solution of the SDE \eqref{eq:sde_m32} -- can be identified as a stationary Gaussian process with a Mat\'ern (index 3/2) covariance function and a PSD given by
\begin{equation}
\label{eq:psd_matern32} H(f) = \dfrac{ 24 \sqrt{3 }\sigma^2 l }{ \left( 3 + (2\pi f l)^2 \right)^2 } \,.
\end{equation}
The parameters $\sigma$ and $l$ can be related to GWB parameters, i.e., $\sigma \sim A_{\rm gw}^\star$ and $l^{-1} \sim f^\star$ where $f^\star$ is a frequency scale such that for $f \ll f^\star$ the PSD is flat ($H(f)\sim 8 \sqrt{3} (A_{\rm gw}^\star)^2 / (3 f^\star)$) and for $f \gg f^\star$ the PSD approaches a power law $H(f) \sim \left(3\sqrt{3}(A_{\rm gw}^\star)^2 / (2 \pi^4 f^\star)\right) ({f/f^\star})^{-4}$. In contrast with the OU process PSD \eqref{eq:psd_exp_kernel}, the PSD \eqref{eq:psd_matern32} of the Mat\'ern process is sufficiently steep to accommodate the traditional interpretation of a GWB due to circular SMBHBs $H(f) \sim f^{-7/3}$. In addition, we shall find that the timing residuals produced by the Brownian harmonic oscillator are also stationary.

We also want to derive PDFs, as we did with the OU process. To avoid repetition, we simply state the results here (utilizing Appendix \ref{sec:a_useful_lemma_on_random_walks}). The likelihood of pulsar redshifts is given by
\begin{equation}
\begin{split}
    & \ln P(z_a, t, z_b, t') \phantom{\dfrac{1}{1}} \\
    & \,\,\,\, = -\dfrac{1}{2} \ln | 2\pi C^Z_{ab}(t,t') | - \dfrac{1}{2} z_a C^Z_{ab}(t,t')^{-1} z_b \,,
\end{split}
\end{equation}
where the covariance
is given by
\begin{align}
C^Z_{ab}(t,t') = & \, \Gamma_{ab} \sigma^2 \bigg[ e^{-\sqrt{3}|t'-t|/l} \left( 1 + \sqrt{3}\dfrac{|t'-t|}{l} \right) \nonumber \\
& - e^{-\sqrt{3}(t'+t)/l} \bigg( 1 + 6 \dfrac{tt'}{l^2} + \sqrt{3}\dfrac{t'+t}{l} \bigg) \bigg]  \,.
\end{align}
At large times compared to the initial time, we obtain
\begin{equation}
C^Z_{ab}(t,t') \sim \Gamma_{ab} \sigma^2 e^{-\sqrt{3}|t'-t|/l} \left( 1 + \sqrt{3}\dfrac{|t'-t|}{l} \right) \,.
\end{equation}
The result clearly reduces to the stationary Gaussian likelihood of pulsar redshifts $z_a$ at time $t$ and $z_b$ at time $t'$, determined by a time-domain covariance function with a corresponding PSD \eqref{eq:psd_matern32}.
This supports the equivalent description of a random phenomena of a Brownian harmonic oscillator and a stationary Gaussian process based on the Mat\'ern index 3/2 kernel.

The PDF of the pulsar timing residuals can be obtained similarly; recall \eqref{eq:redshift_to_residual}. The relevant solution with $r_{a,0}=0$ is given by\footnote{The double integral has been converted into a single integral by using partial integration with the identity
\begin{equation}
\nonumber
\begin{split}
& \int_0^{t_2} dt_1 \, e^{-\sqrt{3}(t_2-t_1)/l} (t_2-t_1) w_a(t_1) \\
& = \dfrac{l^2}{3} w_a(t_2) - \dfrac{d}{dt_2} \left[\dfrac{l}{\sqrt{3}} \int_0^{t_2} dt_1 \, e^{-\sqrt{3}(t_2-t_1)/l} \left( t_2 - t_1 + \dfrac{l}{\sqrt{3}} \right) w_a(t_1)  \right]  \,.
\end{split}
\end{equation}
}
\begin{equation}
    r_a(t) = -\dfrac{l}{\sqrt{3}} \int_0^t dt_1 \, e^{-\sqrt{3}(t-t_1)/l} \left( t - t_1 \right) w_a(t_1) \,.
\end{equation}
This turns out to be $r_a(t)=-l z_a(t)/\sqrt{3}$. The covariance of the pulsar timing residuals can be deduced to be $C^R_{ab}(t,t') = (l^2/3) C^Z_{ab}(t,t')$. In contrast with the OU process, the covariance of the timing residuals does not grow linearly in time; this is because the restoring force or the harmonic potential prevents the particle from wandering off indefinitely in phase space (position and momentum). The PDF of the pulsar timing residuals can be written down straightforwardly.

We note that the SDE \eqref{eq:sde_m32} is obtained as the
over‑damped limit of the full Brownian harmonic oscillator \eqref{eq:langevin_harmonic}.
The underdamped and critically‑damped regimes can be treated analogously, leading to
covariance functions and PSDs that differ from the Mat\'ern-3/2
form \eqref{eq:psd_matern32} but retain the same high frequency steepness \cite{2017AJ....154..220F}. For a
detailed derivation the reader may consult the timeless treatments of the damped
harmonic oscillator of
\cite{Uhlenbeck:1930zz, Chandrasekhar:1943ws}. Because the equation of motion
\eqref{eq:langevin_harmonic} is linear and second order, the PSD of the associated
redshift (velocity) is flat at low frequencies and scales as $f^{-4}$ at high
frequencies, while the timing‑residual (position) PSD scales as $f^{-2}$ to
$f^{-6}$ \cite{Cai:2020ovp, Niu:2021nic, Cai:2023ykr}. If an even steeper spectral slope is required, one must either consider
higher‑order linear SDEs (see Appendix \ref{sec:GPs_and_SDEs}) or introduce higher‑order
restoring forces.

The foregoing results may suggest that the damped harmonic oscillator is mathematically
simpler than the free Brownian particle.  The solution presented in
Sec.~\ref{subsec:free_brownian_motion} is valid for arbitrary initial conditions,
whereas for the oscillator we have restricted ourselves to a special choice in order to highlight the long time steady‑state behavior of the system.
This restriction does not affect the long‑time solution, which is stationary, independent of the
initial state and is the quantity of primary interest in pulsar timing.
Extending the analysis to generic initial conditions is straightforward and is left
as an exercise for the reader.

\subsection{Intrinsic pulsar noise}
\label{subsec:intrinsic_pulsar_noise}

We consider a two-component neutron star model \cite{1969Natur.224..872B} with spin wandering \cite{Meyers:2021myb, Meyers:2021slh, ONeill:2024uiw, Dong:2025nho} to describe pulsar timing noise (Section \ref{subsec:correlated_noises}). For brevity, pulsar subscripts $a, b$ will be suppressed in this section; noting that timing noises are uncorrelated across pulsars. The covariances obtained can be readily generalized to multiple pulsars by multiplying with Kronecker deltas $\delta_{ab}$; in place of the HD correlation $\Gamma_{ab}$ for a GW signal.

The neutron star two-component model with spin wandering can be described by the following SDE \cite{1969Natur.224..872B, Meyers:2021myb, Meyers:2021slh}:
\begin{eqnarray}
\label{eq:crust_eom}
    I_{\rm c} \dfrac{d \Omega_{\rm c} }{dt} &=& N_{\rm c} + \xi_{\rm c}(t) - \dfrac{I_{\rm c}}{\tau_{\rm c}} \left( \Omega_{\rm c} - \Omega_{\rm s} \right)  \\
\label{eq:superfluid_eom}
    I_{\rm s} \dfrac{d \Omega_{\rm s} }{dt} &=& N_s + \xi_{\rm s}(t) + \dfrac{I_{\rm s}}{\tau_{\rm s}} \left( \Omega_{\rm c} - \Omega_{\rm s} \right) \,,
\end{eqnarray}
where $\Omega_{\rm c}$ and $\Omega_{\rm s}$ are the angular velocities of the crust and superfluid components respectively, $I_{\rm c}$ and $I_{\rm s}$ are the corresponding moments of inertia, $N_{\rm c}$ and $N_{\rm s}$ are external torques acting on each component, $\tau_{\rm c}$ and $\tau_{\rm s}$ are the internal coupling time scales, and $\xi_{\rm c}(t)$ and $\xi_{\rm s}(t)$ are stochastic torques acting on each component satisfying
\begin{eqnarray}
\label{eq:xi_c_mean}    \langle \xi_{\rm c}(t) \rangle &=& 0 \,, \\
\label{eq:xi_c_cov}    \langle \xi_{\rm c}(t) \xi_{\rm c}(t') \rangle &=& \sigma_{\rm c}^2 \delta(t - t') \,, \\
\label{eq:xi_s_mean}    \langle \xi_{\rm s}(t) \rangle &=& 0 \,, \\
\label{eq:xi_s_cov}    \langle \xi_{\rm s}(t) \xi_{\rm s}(t') \rangle &=& \sigma_{\rm s}^2 \delta(t - t') \,, \\
\label{eq:cross_xi_c_xi_s}    \langle \xi_{\rm c}(t) \xi_{\rm s}(t') \rangle &=& 0 \,.
\end{eqnarray}
Following \cite{Meyers:2021myb, Meyers:2021slh}, we shall treat the external torques $N_{\rm c}$ and $N_{\rm s}$ as constants.
The case $\langle \xi_{\rm c}(t) \xi_{\rm s}(t') \rangle \neq 0$ can be treated should observations deem it necessary. Spin wandering is therefore associated with the stochastic torques. The last terms in the right hand sides of (\ref{eq:crust_eom}-\ref{eq:superfluid_eom}) is mutual friction associated between the interaction of the superfluid vortices to the rigid crust, a.k.a. a vortex-mediated interaction \cite{Meyers:2021myb, Meyers:2021slh}.
The crust angular velocity, $\Omega_{\rm c}(t)$, is related to the pulsar's rotational phase, $\phi_{\rm c}(t)$, via
\begin{equation}
\label{eq:phase_residual_eom}
    \dfrac{d\phi_{\rm c}}{dt} = \Omega_{\rm c}(t) \,.
\end{equation}
The rotational phase can be related to the timing residuals ($r(t) \sim \phi_{\rm c}(t)$) by normalizing with a nominal spin frequency \cite{2025arXiv251011077K, Dong:2025nho}.

A physical way to look at the neutron star two-component model with spin wandering is as a body averaged or coarse-grained limit of a more fundamental two-fluid counterpart \cite{Prix:1999gk, Prix:2001xc, Prix:2002jn}. This assumes a separation between fast and slow oscillating variables in a fluid such that after coarse-graining, or integrating over the fast oscillating modes, the full fluid dynamic description becomes absorbed into constants and variables in (\ref{eq:crust_eom}-\ref{eq:superfluid_eom}). However, this connection is speculative and remains to be shown explicitly.

Analytic solutions to the two-component model with spin wandering have been partially shown in the Appendices of \cite{Meyers:2021myb,Meyers:2021slh}. 
The integral equations for the crust and superfluid angular velocities are shown in \cite{Meyers:2021myb}. The power spectrum of the crust, superfluid and their covariance are shown in \cite{Meyers:2021slh}.

In the following, we provide the full solution in the time-domain.
Defining the variables
\begin{eqnarray}
    \dfrac{1}{\tau}&=& \dfrac{1}{\tau_{\rm c}} + \dfrac{1}{\tau_{\rm s}} \,, \\
    I &=& I_{\rm c} + I_{\rm s} \,, \\
    \tau \Omega_+ &=& \tau_{\rm c} \Omega_{\rm c} + \tau_{\rm s} \Omega_{\rm s} \,, \\
    \Omega_- &=& \Omega_{\rm c} - \Omega_{\rm s} \,,
\end{eqnarray}
then we are able to express the dynamical system as
\begin{eqnarray}
    \dfrac{d \Omega_+ }{dt} &=& \dfrac{N_+}{I}+ \dfrac{\xi_+(t)}{I} \, \\
    \dfrac{d \Omega_- }{dt} &=& -\dfrac{\Omega_-}{\tau}  + \dfrac{N_-}{I} + \dfrac{\xi_-(t)}{I} \,, 
\end{eqnarray}
where
\begin{eqnarray}
    \dfrac{\tau N_+}{I} &=& \dfrac{\tau_{\rm c} N_{\rm c}}{I_{\rm c}} + \dfrac{\tau_{\rm s} N_{\rm s}}{I_{\rm s}} \,, \\
    \dfrac{\tau \xi_+(t)}{I} &=& \dfrac{\tau_{\rm c} \xi_{\rm c}(t)}{I_{\rm c}} + \dfrac{\tau_{\rm s} \xi_{\rm s}(t)}{I_{\rm s}} \,, \\
    \dfrac{N_-}{I} &=& \dfrac{N_{\rm c}}{I_{\rm c}} - \dfrac{N_{\rm s}}{I_{\rm s}} \,, \\
    \dfrac{\xi_-(t)}{I} &=& \dfrac{\xi_{\rm c}(t)}{I_{\rm c}} - \dfrac{\xi_{\rm s}(t)}{I_{\rm s}} \,.
\end{eqnarray}
We refer to $\Omega_+$ as the effective angular velocity, $\Omega_-$ as the crust-core spin lag (or simply `lag'), $\tau$ as the effective correlation or composite time scale, and $I$ as the total moment of inertia.
The partially decoupled system dynamics could be read as OU processes for $\Omega_+$ (with an infinite correlation time scale) and $\Omega_-$ with a correlation time scale $\tau$.\footnote{
    We use the term `partially decoupled' to highlight that the modes $\Omega_+$ and $\Omega_-$ behave independently in the absence of noise, but are generally stochastically coupled.
}
The effective noise covariance functions are given by
\begin{eqnarray}
\label{eq:xip_mean}    \langle \xi_+(t) \rangle &=& 0 \,, \\
\label{eq:xim_mean}    \langle \xi_-(t) \rangle &=& 0 \,, \\
\label{eq:xip_cov}   \langle \xi_+(t) \xi_+(t') \rangle &=& Q_+ I^2 \delta(t - t') \,, \\
\label{eq:xim_cov}    \langle \xi_-(t) \xi_-(t') \rangle &=& Q_- I^2 \delta(t - t') \,, \\
\label{eq:xip_xim_cross}    \langle \xi_+(t) \xi_-(t') \rangle &=& Q_\times I^2 \delta(t - t') \,,
\end{eqnarray}
where
\begin{eqnarray}
    Q_{\rm c} &=& \sigma_{\rm c}^2/I_{\rm c}^2 \,, \\
    Q_{\rm s} &=& \sigma_{\rm s}^2/I_{\rm s}^2 \,, \\
    \tau^2 Q_+ &=& \tau_{\rm c}^2 Q_{\rm c} + \tau_{\rm s}^2 Q_{\rm s} \,, \\
    Q_- &=& Q_{\rm c} + Q_{\rm s} \,, \\
    \tau Q_\times &=& \tau_{\rm c} Q_{\rm c} - \tau_{\rm s} Q_{\rm s} \,.
\end{eqnarray}
The correlation between the effective torques $\xi_+$ and $\xi_-$ is not assumed, but followed based on the definitions.
We discuss the solutions of $\Omega_+$ and $\Omega_-$ separately.

For $\Omega_+$, the formal solution with initial condition $\Omega_+(t_0) = \Omega_{+,0}$ is given by
\begin{equation}
    \Omega_+(t) - \Omega_{+,0} - \dfrac{N_+}{I} (t - t_0) =  \dfrac{1}{I} \int_{t_0}^t dt_1 \, \xi_+(t_1) \,.
\end{equation}
Then, it can be shown that
\begin{align}
\langle \Omega_+(t) \rangle = \, & \Omega_{+,0} + \dfrac{N_+}{I} (t - t_0) \,, \\
\langle \Omega_+(t) \Omega_+(t') \rangle = \, & \langle \Omega_+(t) \rangle \langle \Omega_+(t') \rangle \\
& + Q_+ \left( {\rm min}(t,t')-t_0 \right) \,. \nonumber
\end{align}
The covariance of $\Omega_+$ evolves linearly in time; because of the external (deterministic and stochastic) torques acting on the system. The problem is reminiscent of a random walk with independent and identically distributed steps. The likelihood of realizing $\Omega_+$ at time $t$ and $\Omega_+'$ at time $t'$ is given by
\begin{align}
& \ln P(\Omega_+, t, \Omega_+', t'| \Omega_{+,0}, t_0) \phantom{\dfrac{1}{1}} \nonumber \\
& = -\dfrac{1}{2} \ln | 2\pi C^+ (t,t',t_0) | \nonumber \\
& \,\,\,\,\,\,- \dfrac{1}{2} \Delta \Omega_+ (t) C^+ (t,t',t_0)^{-1} \Delta \Omega_+ (t') \,, \label{eq:likelihood_Omega_plus}
\end{align}
where
\begin{equation}
    \Delta \Omega_+ (t) = \Omega_+(t) - \langle \Omega_+(t) \rangle
\end{equation}
and the covariance is given by
\begin{equation}
    C^+ (t,t',t_0) = Q_+ \left( {\rm min}(t,t')-t_0 \right) \,.
\end{equation}

For $\Omega_-$, the formal solution with initial condition $\Omega_-(t_0) = \Omega_{-,0}$ is given by
\begin{equation}
\begin{split}
    \Omega_-(t) & - \Omega_{-,0} e^{-(t-t_0)/\tau} - \dfrac{N_- \tau}{I} \left( 1 - e^{-(t-t_0)/\tau} \right) \\
    & = \dfrac{1}{I} \int_{t_0}^t dt_1 \, e^{-(t - t_1)/\tau} \xi_-(t_1) \,.
\end{split}
\end{equation}
It can be shown that
\begin{equation}
    \langle \Omega_-(t) \rangle = \Omega_{-,0} e^{-(t-t_0)/\tau} + \dfrac{N_- \tau}{I} \left( 1 - e^{-(t-t_0)/\tau} \right) \,,
\end{equation}
and
\begin{align}
\label{eq:omegaminus_cov}
    \langle \Omega_-(t) \Omega_-(t') \rangle = \, & \langle \Omega_-(t) \rangle \langle \Omega_-(t') \rangle \\
& + \dfrac{Q_- \tau}{2} \left( e^{-|t'-t|/\tau} - e^{-(t'+t-2t_0)/\tau} \right) \,. \nonumber
\end{align}
The likelihood of $\Omega_-$ at time $t$ and $\Omega_-'$ at time $t'$ is given by
\begin{align}
& \ln P(\Omega_-, t, \Omega_-', t'| \Omega_{-,0}, t_0) \phantom{\dfrac{1}{1}} \nonumber \\
& = -\dfrac{1}{2} \ln | 2\pi C^- (t,t',t_0) | \nonumber \\
& \,\,\,    - \dfrac{1}{2} \Delta \Omega_- (t) C^- (t,t',t_0)^{-1} \Delta \Omega_- (t') \,, \label{eq:likelihood_Omega_minus}
\end{align}
where
\begin{equation}
    \Delta \Omega_- (t) = \Omega_-(t) - \langle \Omega_-(t) \rangle
\end{equation}
and the covariance is given by
\begin{equation}
    C^- (t,t',t_0) = \dfrac{Q_- \tau}{2} \left( e^{-|t'-t|/\tau} - e^{-(t'+t-2t_0)/\tau} \right) \,.
\end{equation}
At large times compared to the initial time, $t', t \gg t_0$, the lag $\Omega_- = \Omega_{\rm c} - \Omega_{\rm s}$ approaches a stationary random process; with a covariance $C^-(t,t',t_0)\sim \left( Q_- \tau/2 \right) e^{-|t' - t|/\tau}$. This is consistent with a Gaussian process with an exponential kernel of correlation scale $\tau$ and amplitude $Q_- \tau/ 2 $.

Additionally, because the torques $\xi_+$ and $\xi_-$ are correlated, the joint likelihood of $\Omega_+$ and $\Omega_-$ at different times would involve cross covariance terms. It can be shown that the cross covariance is given by
\begin{align}
& \langle \Omega_+(t) \Omega_-(t') \rangle - \langle \Omega_+(t) \rangle \langle \Omega_-(t') \rangle  \nonumber \\
& \,\, = Q_\times \tau \left( e^{-(t'-t)/\tau} \Theta(t' - t) + \Theta(t - t') - e^{-( t'-t_0 )/\tau} \right) \,,
\end{align}
where $\Theta(x)$ is the Heaviside step function.
The result for $\langle \Omega_-(t) \Omega_+(t') \rangle - \langle \Omega_-(t) \rangle \langle \Omega_+(t') \rangle$ can be obtained by swapping $t$ and $t'$ in the right hand side of the last expression. Then, at large times compared to the initial time, or $t' \gtrsim t \gg t_0$, it can be shown that $\langle \Delta \Omega_+(t) \Delta \Omega_-(t') \rangle + \langle \Delta \Omega_-(t) \Delta \Omega_+(t') \rangle \sim Q_\times \tau e^{-|t'-t|/\tau}$, or that the full cross covariance at long times depends only on the time lag. This result shows that the diffusive mode $\Omega_+$ drives the damped mode $\Omega_-$ through their noise coupling to produce an extra stationary contribution to the process covariance. The term vanishes only when the noise coupling drops to zero.

The remaining task is to revert the variables back to the original crust and superfluid angular velocities, $\Omega_{\rm c}$ and $\Omega_{\rm s}$. This can be done straightforwardly by using the relations
\begin{eqnarray}
    \Omega_{\rm c} &=& \dfrac{\tau_{\rm s} }{ \tau_{\rm c} + \tau_{\rm s} } \Omega_- + \dfrac{\tau}{ \tau_{\rm c} + \tau_{\rm s} } \Omega_+ \,, \\
    \Omega_{\rm s} &=& - \dfrac{\tau_{\rm c} }{ \tau_{\rm c} + \tau_{\rm s} } \Omega_- + \dfrac{\tau}{ \tau_{\rm c} + \tau_{\rm s} } \Omega_+ \,.
\end{eqnarray}
The mean values can be calculated;
\begin{equation}
    \langle \Omega_{\rm c}(t) \rangle = \dfrac{\tau_{\rm s} }{ \tau_{\rm c} + \tau_{\rm s} } \langle \Omega_-(t) \rangle + \dfrac{\tau}{ \tau_{\rm c} + \tau_{\rm s} } \langle \Omega_+(t) \rangle \,,
\end{equation}
and
\begin{equation}
    \langle \Omega_{\rm s}(t) \rangle = - \dfrac{\tau_{\rm c} }{ \tau_{\rm c} + \tau_{\rm s} } \langle \Omega_-(t) \rangle + \dfrac{\tau}{ \tau_{\rm c} + \tau_{\rm s} } \langle \Omega_+(t) \rangle \,.
\end{equation}
Recall that at large times compared to the initial time, $\langle \Omega_-(t) \rangle \sim N_- \tau/I$ while $\langle \Omega_+(t) \rangle \sim N_+ t/I$. Then, the mean values $\langle \Omega_{\rm c}(t) \rangle$ and $\langle \Omega_{\rm s}(t) \rangle$ grow linearly in time.

The covariances can be calculated similarly. The covariance of the crust angular velocity becomes
\begin{align}
    & \langle \Omega_{\rm c}(t) \Omega_{\rm c}(t') \rangle \nonumber \\
    & = \dfrac{\tau_{\rm s}^2}{(\tau_{\rm c} + \tau_{\rm s})^2} \langle \Omega_-(t) \Omega_-(t') \rangle \nonumber\\
    & \,\,\,\,\,\,+ \dfrac{\tau_{\rm c} \tau_{\rm s}^2}{(\tau_{\rm c} + \tau_{\rm s})^3} \bigg( \langle \Omega_-(t) \Omega_+(t') \rangle + \langle \Omega_+(t) \Omega_-(t') \rangle \bigg) \nonumber \\
    & \,\,\,\,\,\,+ \dfrac{\tau_{\rm c}^2 \tau_{\rm s}^2}{(\tau_{\rm c} + \tau_{\rm s})^4} \langle \Omega_+(t) \Omega_+(t') \rangle   \,.
\end{align}
The covariance of the superfluid angular velocity becomes
\begin{align}
    & \langle \Omega_{\rm s}(t) \Omega_{\rm s}(t') \rangle \nonumber \\
    & = \dfrac{\tau_{\rm c}^2}{(\tau_{\rm c} + \tau_{\rm s})^2} \langle \Omega_-(t) \Omega_-(t') \rangle \nonumber\\
    & \,\,\,\,\,\,- \dfrac{\tau_{\rm c}^2 \tau_{\rm s}}{(\tau_{\rm c} + \tau_{\rm s})^3} \bigg( \langle \Omega_-(t) \Omega_+(t') \rangle + \langle \Omega_+(t) \Omega_-(t') \rangle \bigg) \nonumber \\
    & \,\,\,\,\,\,+ \dfrac{\tau_{\rm c}^2 \tau_{\rm s}^2}{(\tau_{\rm c} + \tau_{\rm s})^4} \langle \Omega_+(t) \Omega_+(t') \rangle   \,.
\end{align}
The crust-superfluid cross covariance becomes
\begin{align}
    & \langle \Omega_{\rm c}(t) \Omega_{\rm s}(t') \rangle \nonumber \\
    & = -\dfrac{\tau_{\rm c} \tau_{\rm s}}{(\tau_{\rm c} + \tau_{\rm s})^2} \langle \Omega_-(t) \Omega_-(t') \rangle \nonumber\\
    & \,\,\,\,\,\,+ \dfrac{\tau_{\rm c} \tau_{\rm s}}{(\tau_{\rm c} + \tau_{\rm s})^3} \bigg( \tau_{\rm s} \langle \Omega_-(t) \Omega_+(t') \rangle - \tau_{\rm c} \langle \Omega_+(t) \Omega_-(t') \rangle \bigg) \nonumber \\
    & \,\,\,\,\,\,+ \dfrac{\tau_{\rm c}^2 \tau_{\rm s}^2}{(\tau_{\rm c} + \tau_{\rm s})^4} \langle \Omega_+(t) \Omega_+(t') \rangle   \,.
\end{align}
Recall the large time asymptotic forms of the covariances of $\Omega_-$, $\Omega_+$, and their cross covariance; for $t,t' \gg t_0$, we have
\begin{equation}
    \langle \Omega_-(t) \Omega_-(t') \rangle - \langle \Omega_-(t) \rangle \langle \Omega_-(t') \rangle \sim  \dfrac{Q_- \tau}{2} e^{-|t'-t|/\tau} \,,
\end{equation}
\begin{equation}
    \langle \Omega_+(t) \Omega_+(t') \rangle - \langle \Omega_+(t) \rangle \langle \Omega_+(t') \rangle \sim Q_+ {\rm min}(t,t') \,,
\end{equation}
and
\begin{align}
    & \langle \Omega_+(t) \Omega_-(t') \rangle - \langle \Omega_+(t) \rangle \langle \Omega_-(t') \rangle \nonumber \\
    & \,\,\,\,\,\,\,\, \sim Q_\times \tau \left( e^{-(t'-t)/\tau} \Theta(t' - t) + \Theta(t - t') \right) \,.
\end{align}
The corresponding result for $\langle \Omega_-(t) \Omega_+(t') \rangle - \langle \Omega_-(t) \rangle \langle \Omega_+(t') \rangle$ can be obtained by interchanging the roles of $t$ and $t'$ in the right hand side of the last expression.
The large time asymptotic form of the covariances of $\Omega_{+}$ and $\Omega_{-}$ shows that the crust and superfluid angular velocity covariances, and the crust-superfluid cross covariance increases linearly in time at the rate $Q_+$. This characteristic nonstationarity is driven by the absence of dynamical friction in the mode $\Omega_+$. The contribution of the (crust-core lag) mode $\Omega_-$ is stationary and decays exponentially with the time lag $|t' - t|/\tau$.

It is useful to write down explicitly the result for the crust angular velocity mean and covariance at large times compared to the initial time; since this is related to radio timing observations. This is given by
\begin{align}
\label{eq:crust_angular_velocity_steady_state}
    \langle \Omega_{\rm c}(t) \rangle \sim & \dfrac{\tau_{\rm s} \tau }{ \tau_{\rm c} + \tau_{\rm s} } \dfrac{N_-}{I} + \dfrac{\tau t}{ \tau_{\rm c} + \tau_{\rm s} } \dfrac{N_+}{I} \,,
\end{align}
and\footnote{
Note that for an arbitrary function $f(x)$, we have
\begin{equation}
    f(|x|) = \Theta(x) f(x) + \Theta(-x) f(-x) \,.
\end{equation}
}
\begin{align}
\label{eq:crust_angular_velocity_cov_steady_state}
    & \langle \Omega_{\rm c}(t) \Omega_{\rm c}(t') \rangle - \langle \Omega_{\rm c}(t) \rangle \langle \Omega_{\rm c}(t') \rangle \nonumber \\
    & \sim \left[ \dfrac{Q_- \tau}{2} \dfrac{\tau_{\rm s}^2}{(\tau_{\rm c} + \tau_{\rm s})^2} + Q_\times \tau \dfrac{\tau_{\rm c}\tau_{\rm s}^2}{(\tau_{\rm c} + \tau_{\rm s})^3}  \right] e^{-|t'-t|/\tau} \nonumber \\
    & \,\,\,\,\,\,\,\, + \dfrac{\tau_{\rm c} \tau_{\rm s}^2}{(\tau_{\rm c} + \tau_{\rm s})^3} Q_\times \tau + \dfrac{\tau_{\rm c}^2 \tau_{\rm s}^2 }{ (\tau_{\rm c} + \tau_{\rm s})^4} Q_+ {\rm min}(t,t') \,.
\end{align}
Consequently, $t=t'$ limit of the above covariance, or the mean square displacement of the crust angular velocity, evolves linearly in observation time because of the absence of dynamical friction in the mode $\Omega_+$. The mean lag and acceleration are also useful to express explicitly in terms of the crust-superfluid parameters;
\begin{equation}
\label{eq:crust_acceleration_steady_state}
    \dfrac{d}{dt} \langle \Omega_{\rm c}(t) \rangle \sim \dfrac{1}{\tau_{\rm c} + \tau_{\rm s}} \left( \dfrac{\tau_{\rm c} N_{\rm c}}{I_{\rm c}} + \dfrac{\tau_{\rm s} N_{\rm s}}{I_{\rm s}} \right) \,
\end{equation}
and
\begin{equation}
\label{eq:lag_steady_state}
    \langle \Omega_{\rm c}(t) - \Omega_{\rm s}(t) \rangle \sim \tau \left( \dfrac{N_{\rm c}}{I_{\rm c}} - \dfrac{N_{\rm s}}{I_{\rm s}} \right) \,.
\end{equation}
It can also be confirmed that $d \langle \Omega_{\rm s} \rangle /dt \sim d \langle \Omega_{\rm c} \rangle /dt $, or that the crust and superfluid will accelerate at the same rate on average. These agree with \cite{Meyers:2021myb, Meyers:2021slh, ONeill:2024uiw}. The covariance of the above parameters can be shown to be given by\footnote{The author thanks Nicholas O'Neill for spotting a missing contribution of the crust-core lag in the covariance of the crust angular acceleration in an earlier version of the manuscript. This term was inadvertently dropped in the text and was not used in any simulations or numerical results.}
\begin{align}
\label{eq:crust_acceleration_cov_steady_state}
    \langle \dot{\Omega}_{\rm c}(t) \dot{\Omega}_{\rm c}(t')  \rangle \sim & \, \langle \dot{\Omega}_{\rm c}(t) \rangle \langle \dot{\Omega}_{\rm c}(t')  \rangle \nonumber \\
    & \,\,\,\, + \left[ (Q_{\rm c} + Q_{\rm s}) \dfrac{\tau}{2 \tau_{\rm c}^2} - \dfrac{Q_{\rm c}}{\tau_{\rm c}} \right] e^{-|t'-t|/\tau} \nonumber \\
    & \,\,\,\, + Q_{\rm c} \delta(t-t')
\end{align}
and
\begin{align}
\label{eq:lag_cov_steady_state}
    & \langle \left( \Omega_{\rm c}(t)-\Omega_{\rm s}(t) \right) \left( \Omega_{\rm c}(t')-\Omega_{\rm s}(t') \right) \rangle \nonumber \phantom{\dfrac{1}{1}} \\
    & \,\,\,\, \sim \langle \left( \Omega_{\rm c}(t)-\Omega_{\rm s}(t) \right) \rangle \langle \left( \Omega_{\rm c}(t')-\Omega_{\rm s}(t') \right) \rangle \nonumber \phantom{\dfrac{1}{1}} \\
    & \,\,\,\,\,\,\,\,\,\,\,\, + (Q_{\rm c} + Q_{\rm s}) \dfrac{\tau}{2} e^{-|t'-t|/\tau} \,.
\end{align}
The expression for the crust angular acceleration covariance can be derived starting with \eqref{eq:crust_eom}. Then, it can be recognized that the crust acceleration variable is a sum of two Gaussian random variables, the crust-core lag and the stochastic torque acting on the crust.
Squaring \eqref{eq:crust_eom}, and evaluating the ensemble average of the resulting expression, utilizing \eqref{eq:xi_c_cov}, \eqref{eq:omegaminus_cov}, and the analytical expression $\langle \Omega_-(t)\xi_{\rm c}(t') \rangle+\langle \xi_{\rm c}(t) \Omega_-(t') \rangle = I_{\rm c} Q_{\rm c} e^{-|t'-t|/\tau}$ leads to \eqref{eq:crust_acceleration_cov_steady_state}. The exact analytical expression for the crust-core lag covariance is given by \eqref{eq:omegaminus_cov}.

In place of the original eight $\tau_{\rm c}$, $\tau_{\rm s}$, $I_{\rm c}$, $I_{\rm s}$, $\sigma_{\rm c}$, $\sigma_{\rm s}$, $N_{\rm c}$, and $N_{\rm s}$, six independent parameters were shown to be potentially constrainable in \cite{ONeill:2024uiw, Dong:2025nho}: $\tau = \tau_{\rm c} \tau_{\rm s}/(\tau_{\rm c} + \tau_{\rm s})$, $r= \tau_{\rm s}/\tau_{\rm c}$, $Q_{\rm c}=\sigma_{\rm c}^2/I_{\rm c}^2$, $Q_{\rm s}=\sigma_{\rm s}^2/I_{\rm s}^2$, $\llangle \dot{\Omega}_{\rm c} \rrangle$, and $\llangle \Omega_{\rm c}-\Omega_{\rm s} \rrangle$; where $\llangle \dot{\Omega}_{\rm c} \rrangle$, and $\llangle \Omega_{\rm c}-\Omega_{\rm s} \rrangle$ are the constant steady state limits of the crust angular acceleration and the crust-core lag, given by (\ref{eq:crust_acceleration_steady_state}-\ref{eq:lag_steady_state}). The inverse relations are given by $\tau_{\rm c} = (1 + r^{-1}) \tau$, $\tau_{\rm s} = (1 + r) \tau$, $N_{\rm c}/I_{\rm c} = \llangle \dot{\Omega}_{\rm c} \rrangle + \llangle \Omega_{\rm c}-\Omega_{\rm s} \rrangle r / (\tau (1 + r))$, and $N_{\rm s}/I_{\rm s} = \llangle \dot{\Omega}_{\rm c} \rrangle - \llangle \Omega_{\rm c}-\Omega_{\rm s} \rrangle / (\tau (1 + r))$; expressing the parameters of the two-component model using the crust-superfluid coupling time scale, crust-superfluid time scale ratio, the pulsar spin down rate, and the crust-core lag.
In \cite{ONeill:2024uiw}, the parameters $\tau$, $Q_{\rm c}=\sigma_{\rm c}^2/I_{\rm c}^2$, $Q_{\rm s}=\sigma_{\rm s}^2/I_{\rm s}^2$, and $\langle \dot{\Omega}_{\rm c} \rangle$ were constrained using PSR J1359-6038; the parameters $r= \tau_{\rm s}/\tau_{\rm c}$ and the lag $\langle \Omega_{\rm c}-\Omega_{\rm s}\rangle $ remained unconstrained.

The system behavior in certain limits can be illuminating. For $N_{\rm c}/I_{\rm c} \ll N_{\rm s}/I_{\rm s} \leq 0$, the crust acceleration and the crust-core spin lag become $ \langle \dot{\Omega}_{\rm c} \rangle \sim \tau N_{\rm c}/( \tau_{\rm s} I_{\rm c} ) \leq 0$ and $\langle \Omega_{\rm c}-\Omega_{\rm s} \rangle \sim \tau_{\rm s} \langle \dot{\Omega}_{\rm c} \rangle \leq 0$. For $N_{\rm s}/I_{\rm s} \ll N_{\rm c}/I_{\rm c} \leq 0$, the crust acceleration and the lag becomes $ \langle \dot{\Omega}_{\rm c} \rangle \sim \tau N_{\rm s}/( \tau_{\rm c} I_{\rm s} ) \leq 0$ and $\langle \Omega_{\rm c}-\Omega_{\rm s} \rangle \sim -\tau_{\rm c} \langle \dot{\Omega}_{\rm c} \rangle \geq 0$. Negative torques naturally support the spin down of the crust, $\langle \dot{\Omega}_{\rm c} \rangle < 0$. The crust-core lag is positive when the superfluid torque dominates, and negative when the crust torque dominates. These limits can be utilized to build priors on the steady state parameters; such as
\begin{align}
    \langle \dot{\Omega}_{\rm c} \rangle_{\rm min} \leq \langle \dot{\Omega}_{\rm c} \rangle < 0
\end{align}
and
\begin{align}
    & \tau_{\rm max}( 1 + r_{\rm max} ) \langle \dot{\Omega}_{\rm c} \rangle_{\rm min} \nonumber \\
    & \,\,\,\,\,\,\,\, \leq \langle \Omega_{\rm c}-\Omega_{\rm s} \rangle \leq - \tau_{\rm max} \left( 1 + r_{\rm min}^{-1} \right) \langle \dot{\Omega}_{\rm c} \rangle_{\rm min} \,.
\end{align}
The prior on the lag considered in \cite{ONeill:2024uiw} when the model was tested in PSR J1359-6038 holds for $r = 1$ ($\tau_{\rm c} = \tau_{\rm s}$). The $r$ factors in the upper and lower limits are negligible when $r$ is ${\cal O}(1)$. However, for the case in \cite{ONeill:2024uiw}, $\tau_{\rm max} \langle \dot{\Omega}_{\rm c} \rangle_{\rm min} = - 10^{-3} \, {\rm rad} \, {\rm s}^{-1}$ and $10^{-2} \leq r \leq 10^2$ (based on models of stellar structure \cite{Link:1999ca, 2000MNRAS.315..534L, 2011MNRAS.414.1679E, Chamel:2012zn}), and so factoring in the $r$ dependence in the lag prior results in a range that is two orders of magnitude larger compared to $ | \tau_{\rm max} \langle \dot{\Omega}_{\rm c} \rangle_{\rm min} | $, i.e, $-10^{-1} \, {\rm rad} \, {\rm s}^{-1} \leq \langle \Omega_{\rm c}-\Omega_{\rm s} \rangle \leq 10^{-1} \, {\rm rad} \, {\rm s}^{-1}$.

The cross covariance between the crust and superfluid angular velocities at large times might turn out to be useful;
\begin{align}
    & \langle \Omega_{\rm c}(t) \Omega_{\rm s}(t') \rangle - \langle \Omega_{\rm c}(t) \rangle \langle \Omega_{\rm s}(t') \rangle \nonumber \\
    & \sim - \dfrac{\tau_{\rm c} \tau_{\rm s}}{(\tau_{\rm c} + \tau_{\rm s})^2} \dfrac{Q_- \tau}{2} e^{-|t'-t|/\tau} + \dfrac{\tau_{\rm c}^2 \tau_{\rm s}^2 }{ (\tau_{\rm c} + \tau_{\rm s})^4} Q_+ {\rm min}(t,t') \nonumber \\
    & \,\,\,\,\,\,\,\, + \dfrac{\tau_{\rm c} \tau_{\rm s} }{ (\tau_{\rm c} + \tau_{\rm s})^3} Q_\times \tau \bigg[ \Theta(t' - t)\left( \tau_{\rm s} - \tau_{\rm c} e^{-(t'-t)/\tau} \right)   \nonumber \\
    & \phantom{ggggggggggggggiii} + \Theta(t - t')\left( - \tau_{\rm c} + \tau_{\rm s} e^{-(t-t')/\tau} \right) \bigg] \,.
\end{align}

\begin{figure*}[t]
\centering
\subfigure[]{
    \includegraphics[width=0.475\textwidth]{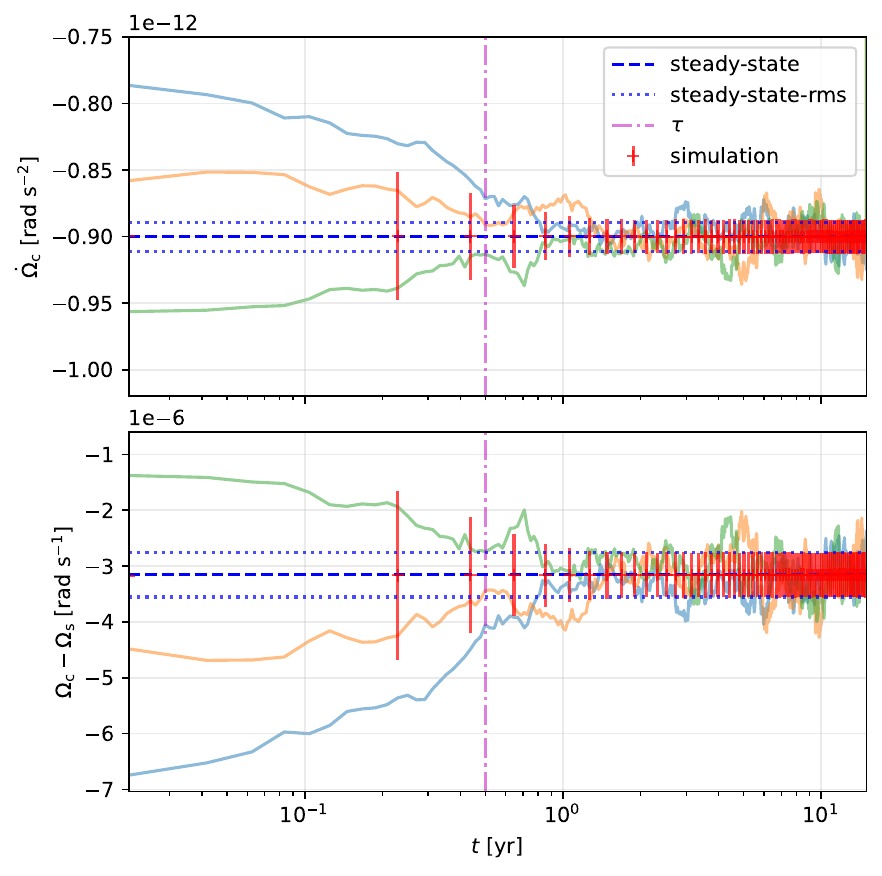}
}
\subfigure[]{
    \includegraphics[width=0.475\textwidth]{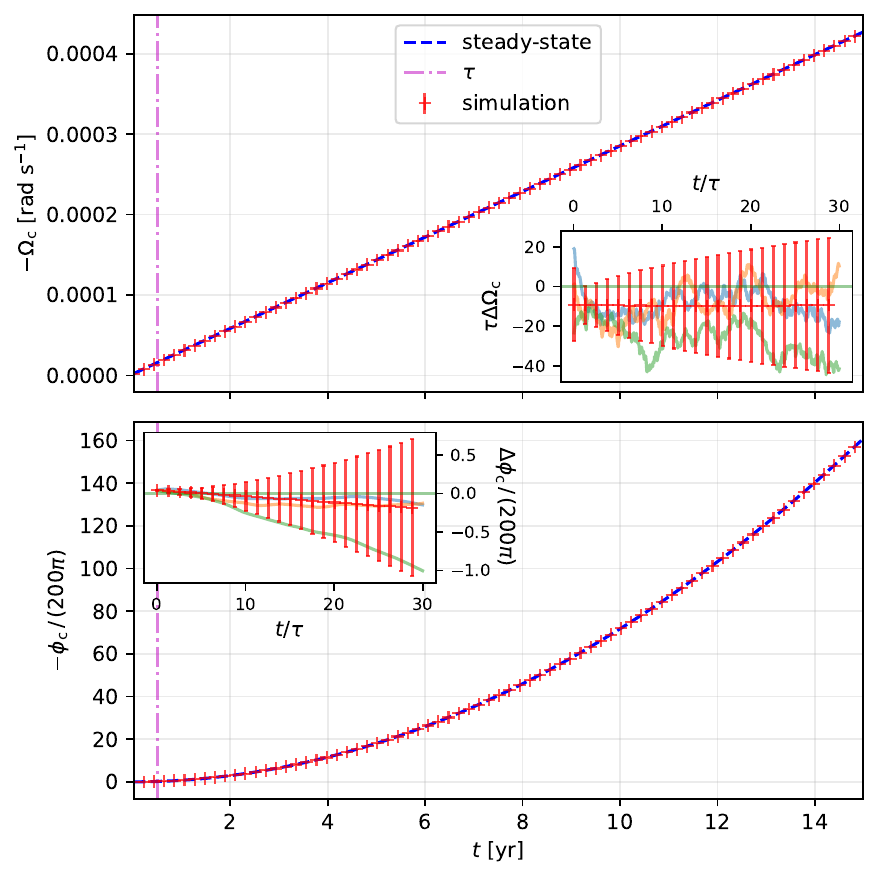}
}
\caption{
Simulation results (red points with error bars) compared with the steady‑state analytical predictions (blue solid line for the mean and blue dotted lines for the rms) for  
(a) the crust angular acceleration and the crust–core lag, and  
(b) the crust angular velocity and the phase.  
The parameters used are  
$\tau_{\rm c}=\tau_{\rm s}=1\,{\rm yr}$,  
$Q_{\rm c}=Q_{\rm s}=10^{-20}\,{\rm rad}^{2}\,{\rm s}^{-3}$,  
$N_{\rm c}/I_{\rm c}=-10^{-12}\,{\rm rad}\,{\rm s}^{-2}$,  
$N_{\rm s}/I_{\rm s}=0.8\,N_{\rm c}/I_{\rm c}$,  
with uniformly drawn initial conditions centered at $\Omega_{{\rm c},0}=\Omega_{{\rm s},0}=\phi_{{\rm c},0}=0$.  
These choices yield negative steady‑state values $\llangle\dot{\Omega}_{\rm c}\rrangle$ and 
$\llangle\Omega_{\rm c}-\Omega_{\rm s}\rrangle$, i.e. a pulsar that is spinning down and whose crust lags the superfluid.  
In panel (b) the inset shows the residuals (simulation compared to theory); the horizontal green line marks zero and the error bars represent the combined standard error of the simulation average and the theoretical prediction.  
The faint coloured curves correspond to three individual realisations, while the red points are sample averages over $10000$ realisations.}
\label{fig:two_component_model_simulations}
\end{figure*}

The pulsar phase is obtained by integrating the crust angular velocity;
\begin{equation}
    \phi_{\rm c}(t) - \phi_{{\rm c}, 0} = \int_{t_0}^t dt_1 \, \Omega_{\rm c}(t_1) \,.
\end{equation}
This has the formal solution;\footnote{
This can be derived using the following identities;
\begin{align}
    \int_{t_0}^t dt_2 \int_{t_0}^{t_2} dt_1 \, \xi(t_1) = \int_{t_0}^t dt_1 (t - t_1) \xi(t_1) \,,
\end{align}
and
\begin{align}
    & \int_{t_0}^t dt_2 \int_{t_0}^{t_2} dt_1 \, e^{-(t_2 - t_1)/\tau} \xi(t_1) = \tau \int_{t_0}^t dt_1 \, \left[ 1 - e^{-(t-t_1)/\tau} \right] \xi(t_1) \,.
\end{align}
The first one can be proven by changing the order of integration; the second one by performing integration by parts.
}
\begin{align}
    & \phi_{\rm c}(t) - \langle \phi_{\rm c}(t) \rangle \nonumber \\
    & \,\,= \dfrac{\tau}{(\tau_{\rm c} + \tau_{\rm s})} \dfrac{1}{I} \int_{t_0}^t dt_1 \, \nonumber \\
    & \,\,\,\,\,\,\,\,\, \bigg[ \xi_+(t_1) (t-t_1) + \xi_-(t_1) \tau_{\rm s} \left( 1 - e^{-(t-t_1)/\tau} \right)  \bigg] \,, \label{eq:phase_residual_formal_solution}
\end{align}
where the mean phase residual is given by
\begin{align}
\langle \phi_{\rm c}(t) \rangle = \, \phi_{{\rm c}, 0} & + \dfrac{\tau_{\rm s} }{ \tau_{\rm c} + \tau_{\rm s} } \bigg[ \dfrac{N_- \tau}{I}(t-t_0) \nonumber \\
& + \left( \Omega_{-,0} - \dfrac{N_- \tau}{I} \right) \tau \left( 1 - e^{-(t-t_0)/\tau} \right) \bigg] \nonumber \\
& + \dfrac{\tau}{\tau_{\rm c} + \tau_{\rm s}} \bigg[ \Omega_{+,0}(t-t_0) + \dfrac{N_+}{2I}(t-t_0)^2 \bigg] \,.
\end{align}
The phase residual grows quadratically in time because of the external torques acting on the system. In terms of the steady-state parameters (\ref{eq:crust_acceleration_steady_state}) and (\ref{eq:lag_steady_state}), the mean phase residual at large times compared to the initial time becomes
\begin{align}
\label{eq:phase_residual_steady_state}
\langle \phi_{\rm c}(t) \rangle \sim \dfrac{\tau_{\rm s}}{\tau_{\rm c} + \tau_{\rm s}} (t - \tau) \llangle \Omega_{\rm c} - \Omega_{\rm s} \rrangle + \dfrac{\llangle \dot{\Omega}_{\rm c} \rrangle}{2} t^2 \,.
\end{align}
We have omitted terms that depend on initial conditions such as $\Omega_{+, 0} (t - t_0)$, and have used the notation $\llangle \cdots \rrangle$ to denote constant steady-state values. The constant, linear, and quadratic deterministic trends in time in the phase can be related to the torques acting on the crust and the superfluid.
The covariance of the phase can be shown to be
\begin{align}
& \langle \phi_{\rm c}(t) \phi_{\rm c}(t') \rangle - \langle \phi_{\rm c}(t) \rangle \langle \phi_{\rm c}(t') \rangle \nonumber  \\
& = \dfrac{\tau^2}{(\tau_{\rm c} + \tau_{\rm s})^2} \left[ Q_+ {\cal I}_1(t, t') + Q_\times \tau_{\rm s} {\cal I}_2(t, t') + Q_- \tau_{\rm s}^2 {\cal I}_3(t, t') \right] \,, \label{eq:phase_residual_cov}
\end{align}
where the functions ${\cal I}_1$, ${\cal I}_2$, and ${\cal I}_3$ are given by
\begin{align}
    {\cal I}_1(t, t') = & -\dfrac{1}{6} ( {\rm min}(t,t')-t_0 )^2 \nonumber \phantom{\dfrac{1}{1}} \\
    & \left( ({\rm min}(t,t')-t_0) - 3 ({\rm max}(t,t')-t_0) \right) \,, \label{eq:I1_def} \phantom{\dfrac{1}{1}}
\end{align}
\begin{align}
    {\cal I}_2(t,t') = & - \tau^2 e^{-|t'-t|/\tau} + \tau (t - t_0 + \tau) e^{-(t'-t_0)/\tau} \nonumber \phantom{\dfrac{1}{1}} \\
    & + \tau (t' - t_0 + \tau) e^{-(t - t_0)/\tau} \nonumber \phantom{\dfrac{1}{1}} \\
    & + ({\rm min}(t,t')- t_0 - \tau)({\rm max}(t,t')-t_0 + \tau) \,, \label{eq:I2_def} \phantom{\dfrac{1}{1}}
\end{align}
\begin{align}
    {\cal I}_3(t,t') = & - \dfrac{\tau}{2} e^{-|t'-t|/\tau} - \dfrac{\tau}{2} e^{-(t' + t - 2 t_0)/\tau} \nonumber \phantom{\dfrac{1}{1}}  \\
    & + \tau e^{-(t-t_0)/\tau} + \tau e^{-(t'-t_0)/\tau} \nonumber \phantom{\dfrac{1}{1}} \\
    & + {\rm min}(t,t')- t_0 - \tau \,. \label{eq:I3_def} \phantom{\dfrac{1}{1}}
\end{align}
Each of the terms capture a distinct physical contribution: ${\cal I}_1$ represents the $\Omega_+$ contribution, ${\cal I}_2$ the $\Omega_+$-$\Omega_-$ cross contribution, and ${\cal I}_3$ the $\Omega_-$ contribution.
The detailed derivation of the covariance of the phase is given in Appendix \ref{sec:cov_phase_2cm}.

It is insightful to look at the large time asymptotic limit, $t', t \gg t_0$, of the analytical expressions. Then, the covariance of the phase reduces to
\begin{align}
& \langle \phi_{\rm c}(t) \phi_{\rm c}(t') \rangle - \langle \phi_{\rm c}(t) \rangle \langle \phi_{\rm c}(t') \rangle \nonumber \phantom{\dfrac{1}{1}} \\
& \sim \dfrac{\tau^2}{(\tau_{\rm c} + \tau_{\rm s})^2} \bigg[  \left( \dfrac{Q_+}{6} t^2(3t' -t) + Q_- \tau_{\rm s}^2 t \right) \Theta(t'-t) \nonumber \\
& \phantom{ggggggggiii} + \left( \dfrac{Q_+}{6} t'^2(3t -t') + Q_- \tau_{\rm s}^2 t' \right) \Theta(t-t') \nonumber \\
& \phantom{gggg}  + Q_\times \tau_{\rm s} t t' -\left( Q_\times \tau + \dfrac{Q_- \tau_{\rm s}}{2} \right) \tau_{\rm s} \tau e^{-|t'-t|/\tau}   \bigg] \,. \label{eq:phase_residual_cov_steady_state}
\end{align}
This shows that the system is nonstationary, as expected, because of the torques acting on the system; the deterministic torques contribute nonstationary terms in the means and the stochastic torques to the covariances.
The mean square displacement of the phase, or rather the $t = t'$ limit of the phase covariance, grows cubic in observation time.

Figure \ref{fig:two_component_model_simulations} compares the simulation results with long time or steady-state theoretical expectation values for the mean and root-mean-squares of the crust angular acceleration, lag (\ref{eq:crust_acceleration_steady_state}-\ref{eq:lag_cov_steady_state}), angular velocity (\ref{eq:crust_angular_velocity_steady_state}-\ref{eq:crust_angular_velocity_cov_steady_state}), and phase residual [\eqref{eq:phase_residual_steady_state} and \eqref{eq:phase_residual_cov_steady_state}]. For the simulation, the system (\ref{eq:crust_eom}-\ref{eq:phase_residual_eom}) was nondimensionalized (Appendix \ref{sec:nondimensionalization_of_the_two_component_model}) and solved via the Euler-Maruyama method \cite{1989SHH.....3..155K, 1994ZaMM...74..332V}. The parameters considered are $\tau_{\rm c}=\tau_{\rm s}=1\,{\rm yr}$, $Q_{\rm c}=Q_{\rm s}=10^{-20}\,{\rm rad}^{2}\,{\rm s}^{-3}$, $N_{\rm c}/I_{\rm c}=-10^{-12}\,{\rm rad}\,{\rm s}^{-2}$, and $N_{\rm s}/I_{\rm s}=0.8\,N_{\rm c}/I_{\rm c}$, and the initial conditions given by a uniform distribution of crust and superfluid angular velocities centered at $\Omega_{{\rm c},0}=\Omega_{{\rm s},0}=\phi_{{\rm c},0}=0$. These choices yield a pulsar that is spinning down and whose crust lags the superfluid.
Despite the large initial distribution, the sample averages converge rapidly to the theoretical predictions within a few
composite time scales $\tau$, confirming both the mean trajectories and the
square-root-of-covariance error envelopes.
The time evolution of the means and covariances also directly attests to the nonstationarity of the system.

The physical origin of this nonstationarity can now be stated precisely: it traces to the
coexistence of a diffusive eigenmode $\Omega_+$, which lacks a restoring force and
consequently diffuses without bound, and a damped eigenmode $\Omega_-$, which relaxes
exponentially on the composite time scale $\tau$.
Because the observable pulsar phase is an integral of $\Omega_{\rm c}$, which contains a
contribution from the diffusive mode, its variance grows cubically in time.
This signature cannot be captured by any stationary covariance function.

\section{Discussion}
\label{sec:discussion}

In the preceding section we have shown that SDEs can be treated analytically by utilizing methods that have long been used to describe Brownian motion and diffusion \cite{Uhlenbeck:1930zz, Chandrasekhar:1943ws, 1944BSTJ...23..282R, 1945RvMP...17..323W, 1947AmMM...54..369K}. While any SDE -- no matter how high‑order -- can be integrated numerically, an analytical solution provides direct physical insight and often yields compact expressions that are useful for data analysis pipelines. Analytical solutions have also played a key role in the development of scalable Gaussian processes \cite{1996PhRvE..54.2084G, 2017AJ....154..220F, 2018PhRvE..98a2136S}. In the discussion that follows we will revisit the modelling of the GWB and intrinsic red noise, which are frequently represented by an OU process for the pulsar spin frequency in state space approaches \cite{Kimpson:2025lju, 2025arXiv251011077K}, and examine the two‑component crust-superfluid neutron star model \cite{1969Natur.224..872B, Haskell:2015jra, Meyers:2021myb, Meyers:2021slh} (see Table~\ref{tab:nonstationarity}). Furthermore, we will address two points that have been only glossed over so far: first, how results expressed in terms of ensemble averages can be meaningfully applied to a single system; and second, how the intrinsic nonstationarity of random processes based on SDEs can be handled.

As shown in Section \ref{subsec:free_brownian_motion}, an OU process
applied to the pulsar redshift (or, to the spin frequency\footnote{
The discussion applies equally to redshifts and spin
frequencies because the two quantities are linearly related; the same
statistics therefore hold for the timing and phase residuals, which differ only
by a constant factor given by the nominal spin frequency \cite{Kimpson:2025lju, 2025arXiv251011077K}.}) satisfies the
equation of motion of a free Brownian particle: the redshift plays the role of
the particle’s velocity, while the timing residual corresponds to its position.
Einstein \cite{1905AnP...322..549E} and Langevin \cite{Langevin1908,10.1119/1.18725}
demonstrated that the RMS displacement of a Brownian particle
grows as $\sqrt{t}$, i.e., the variance increases linearly with the elapsed time.
Consequently, the RMS timing residual also grows as $\sqrt{t}$, which is
reflected in the linear‑in‑time term of the covariance
\eqref{eq:timing_residual_cov_freebrownian_steady_state}.  Thus, while the OU
process yields a stationary redshift (its autocorrelation depends only on the
lag $|t-t'|$), the associated integral of the redshift, or the  timing residual, is intrinsically non‑stationary \cite{1996PhRvE..54.2084G}.
The magnitude of this non‑stationarity is set by the observing baseline
$(t-t_{0})$ weighted by the correlation time $\ell$; in the regime $t-t_{0} \gtrsim \ell$ (relevant to PTAs with $t-t_{0}\sim10\,$yr and $\ell\sim1\,$yr), this contribution is nonnegligible. 

\begin{table}[t]
    \renewcommand{\arraystretch}{1.3}
    \centering
    \caption{Random processes solved analytically in this work and their stationarity viewed in terms of pulsar timing observables; pulsar redshifts or timing residuals. OU stands for Ornstein-Uhlenbeck process (free Brownian particle), M32 for Mat\'ern 3/2 (Brownian harmonic oscillator), 2CM to the two-component pulsar timing noise model. OU and M32 apply to both GWB signal (HD correlated) and pulsar red noise. 2CM applies to pulsar timing noise.}
    \begin{tabular}{|c|c|c|c|}
        \hline
        Model & Observable & Stationary & Notes \\ 
        \hline\hline

        \multirow{2}{*}{OU} 
            & Redshifts & {\bf\checkmark} & \multirow{2}{*}{Sec. \ref{subsec:free_brownian_motion}} \\ \cline{2-3}
            & Timing residuals & ${\boldsymbol{ \times }}$ &  \\ 
        \hline

        \multirow{2}{*}{M32} 
            & Redshifts & {\bf\checkmark} & \multirow{2}{*}{Sec. \ref{subsec:brownian_oscillator}} \\ \cline{2-3}
            & Timing residuals & ${\boldsymbol{ \times }}$ & \\
        \hline

        \multirow{2}{*}{2CM} 
            & Frequencies & ${\boldsymbol{ \times }}$ & \multirow{2}{*}{Sec. \ref{subsec:intrinsic_pulsar_noise}} \\ \cline{2-3}
            & Phases & ${\boldsymbol{ \times }}$ & \\
        \hline
    \end{tabular}
    \label{tab:nonstationarity}
\end{table}

It turns out that the nonstationary timing residual generated by an OU pulsar redshift can be partially dealt with by marginalising over a set of deterministic basis functions that capture the long time trends in pulsar timing data \cite{ONeill:2024uiw, 2025arXiv251011077K, Kimpson:2025lju, Dong:2025nho}. In PTA analysis the timing model already includes a constant offset, a linear term, and a quadratic term that accounts for the spin evolution of a pulsar \cite{Pitrou:2024scp, Allen:2025waa}. By projecting the data onto the subspace orthogonal to these basis functions one eliminates the dominant contribution to the linear term that appears in the covariance \eqref{eq:timing_residual_cov_freebrownian_steady_state}.\footnote{\label{footnote:mitigation_of_nonstationarity}
This partial mitigation can also be realized by recognizing that the integral operation on a stationary process $z$ with a PSD $S(f)$ will produce a process $r=\int dt \, z$ with a singular PSD $S(f)/f^2$; the pole at $f=0$ characteristic of nonstationarity. On the other hand, the process ${\cal P}$ of projecting out long term trends in the data suppresses the power below a threshold frequency $f_1=1/T$, where $T$ is the width of the observation window. For constant, linear and quadratic trends (embodied by the quadratic spin down model) in the data, this procedure generates a process ${\cal P}[r]$ with an effective PSD $S(f) {\cal T}(f) / f^2 $ where the transmission function ${\cal T}(f)$ falls off as $\sim f^6$ for $f \lesssim 1/T$ \cite{Hazboun:2019vhv, Pitrou:2024scp, Allen:2025waa}. The suppression of power at low frequencies masks away the pole $f=0$; and the nonstationary behavior associated with it. However, the residuals ${\cal P}[r]$ are also generally nonstationary \cite{Allen:2025waa}.
} In general, the timing model fitting or subtraction of deterministic trends in the timing data make the resulting residuals nonstationary \cite{Lee:2012bf, vanHaasteren:2012hj, Allen:2025waa}. Nonetheless, the resulting residuals can be fairly described by a stationary covariance, which is the quantity normally used in PTA searches for a GWB or red noise processes. This procedure is standard in PTA pipelines and reflects that astrophysical model errors manifest as deterministic trends that can be approximately marginalised over. Consequently, the nonstationarity in the timing residuals in an OU‑modelled pulsar redshift is mitigated in the analysis.

Non‑stationarity is not a flaw of a stochastic model; it is a property that must be
identified and interpreted.  Whether a signal should be treated as stationary or
nonstationary depends on its origin.  
For a GWB generated by a large population of
weak sources \cite{Burke-Spolaor:2018bvk, Domenech:2025ffb} the signal is expected to be a stationary
process (see Section~\ref{subsec:gwb_signal}). In contrast, for a GWB that is
dominated by a handful of bright binaries, the same reasoning to stationarity cannot be applied; the observational implications of
such a scenario are discussed in \cite{Jow:2025uut,2025arXiv251210795T}.  

Because an OU process for the pulsar redshift (or spin
frequency) yields a stationary redshift but a nonstationary timing residual (see also Appendix \ref{sec:generating_matern}),
it is mathematically inconsistent with a truly stationary GWB produced by a
large number of weak sources (Section~\ref{subsec:gwb_signal}). 
Here `mathematically inconsistent' means the timing residual covariance is strictly nonstationary; the nonstationarity is mitigated in practice by projecting out long time trends, although the residuals themselves remain technically nonstationary \cite{Antonelli:2022gqw, Pitrou:2024scp, Allen:2025waa} (see also footnote \ref{footnote:mitigation_of_nonstationarity}). For a truly stationary GWB signal, a more precise, self-consistent stochastic description can be realized by modelling the
redshift with a Mat\'ern‑$3/2$ process, or rather, the overdamped limit of a Brownian
harmonic oscillator (Section~\ref{subsec:brownian_oscillator}).  The harmonic
oscillator introduces a restoring force (Hooke’s law) that confines the random
motion in phase space; consequently both the redshift (velocity) and the timing
residual (position) become stationary, with covariances that depend only on the
time lag $|t-t'|$.  In the OU spin frequency model the restoring force is absent: dynamical
friction damps the velocity but, in the presence of white noise, the position
diffuses without bound, leading to a covariance that grows linearly with the
elapsed time.

Another motivation for modelling spin frequencies or pulsar redshifts with a
Mat\'ern‑$3/2$ process (Section~\ref{subsec:brownian_oscillator}) is that it admits a
single, mathematically well‑defined PSD in the frequency domain that is a Fourier pair of the covariance function. The
OU process has a PSD \eqref{eq:psd_exp_kernel} that scales as $f^{-2}$, i.e., it
is very shallow, whereas the Mat\'ern‑$3/2$ PSD \eqref{eq:psd_matern32}
falls off as $f^{-4}$. The steeper spectral slope of the Mat\'ern kernel
provides greater flexibility when fitting PTA data, because it can accommodate
both the canonical $f^{-7/3}$ redshift PSD of a GWB due to circular binaries and a greater variation in the red noise spectra in pulsars.
This spectral flexibility makes the Mat\'ern-3/2 spin frequency model a more versatile basis for PTA analyses compared to the shallower OU process.

The two‑component neutron‑star model was originally proposed to explain
pulsar glitches \cite{1969Natur.224..872B, Haskell:2015jra} and, more than
five decades later, with a notion of spin wandering, has been demonstrated to provide a statistically consistent
description of pulsar timing noise compared with a one‑component model
\cite{ONeill:2024uiw, Dong:2025nho}.  In Sec.~\ref{subsec:intrinsic_pulsar_noise}
we derived analytical expressions for the means and covariances of the crust
and superfluid angular velocities as well as for the observable pulsar phase.
These results reveal that the system is intrinsically nonstationary \cite{Antonelli:2022gqw, Antonelli:2025vhw}. 
Deterministic torques dictate the evolution of the mean values and stochastic torques
drive the growth of the covariances. An intuitive analogy is a particle in a constant gravitational field, whose position evolves quadratically in time; in a neutron star context the gravitational field is played by the constant torques, the velocity by the spin frequencies, and the position by the rotational phase. The particle also experiences a stochastic driving force; accordingly, the stochastic torques in the two-component model with spin wandering play the role of the random kicks that drive the diffusion. Because the resulting processes are nonstationary, they cannot be represented by a PSD via the Wiener-Khinchin theorem.

Without the deterministic torques, the two-component model with spin wandering can be also approached as a two-dimensional limit of a multivariate OU process. However, the characteristic feature of the model is that the drift or normalized dynamical friction matrix associated with the process is singular, or that one of its eigenvalues is zero. This implies that one of the eigenmodes of the system is diffusive or that it reaches a stationary state only after an infinite time. This singular feature prevents the straightforward application of well-known analytical solutions of the multivariate OU process to the two-component model with spin wandering \cite{1360855570504742656, 2018PhRvE..98a2136S}. This warrants the model an independent treatment. As a matter of fact, the eigenmodes are exactly $\Omega_-$ and $\Omega_+$: $\Omega_-$ is an OU process with a correlation time scale given by the composite time scale, and $\Omega_+$ is a diffusive process. The crust-core lag $\Omega_-\equiv \Omega_{\rm c} - \Omega_{\rm s}$ is a stationary process for this reason. The nonstationarity in the covariance of the model is due to the diffusive eigenmode $\Omega_+ \equiv (\tau_{\rm c} \Omega_{\rm c} + \tau_{\rm s} \Omega_{\rm s})/\tau$. Since the crust and superfluid rotational states are linear combinations of the two eigenmodes, they manifest stationarity due to $\Omega_-$ but also the nonstationarity of $\Omega_+$.

In practice, the nonstationarity in the timing residuals model is mitigated by projecting out, or marginalising over, the low-order deterministic trends (constant, linear and quadratic terms) that are already included in standard PTA analysis \cite{ONeill:2024uiw, Kimpson:2025lju, 2025arXiv251011077K, Dong:2025nho}. After removal of these trends the residuals are described approximately by stationary covariances. It is worth noting that this standard procedure technically produces weakly nonstationary timing residuals \cite{Lee:2012bf, vanHaasteren:2012hj, Allen:2025waa}. Our analytical solutions have been validated against numerical simulations, reproducing both the expected mean trajectories and the associated standard errors (i.e., the covariances).

\begin{table}[t]
\caption{SDE-based applications to radio pulsar timing; SN stands for spin noise; WP stands for Wiener process; OU for Ornstein--Uhlenbeck process; and 2CM for two-component model.}
\renewcommand{\arraystretch}{1.5}
\begin{tabular}{|c|c|c|c|}
\hline
Application & SDE Model & Subject & Data \\
\hline\hline
\multirow{2}{*}{Glitch detection}
  & WP~\cite{Melatos:2020rlu} & \multirow{2}{*}{SN} & Real+Sim. \\
  & WP~\cite{Dunn:2023uqo}    &                             & Real+Sim. \\
\hline
\multirow{4}{*}{Pulsar physics}
  & 2CM~\cite{Meyers:2021myb}    & \multirow{4}{*}{SN} & Simulated \\
  & 2CM~\cite{Meyers:2021slh}    &                             & Simulated \\
  & 2CM+WP~\cite{ONeill:2024uiw} &                             & Real+Sim. \\
  & 2CM~\cite{Dong:2025nho}      &                             & Real \\
\hline
\multirow{2}{*}{\shortstack{CW in PTA}}
  & OU~\cite{Kimpson:2024fel} & \multirow{2}{*}{SN} & Simulated \\
  & OU~\cite{Kimpson:2024kgq} &                             & Simulated \\
\hline
SGWB in PTA
  & OU~\cite{Kimpson:2025lju} & SN+GWB & Simulated \\
\hline
\end{tabular}
\label{tab:sde_radio_timing}
\end{table}

Table \ref{tab:sde_radio_timing} enumerates recent applications of SDEs in radio pulsar timing. Solutions to SDEs are relevant to state space methods which have gained traction in pulsar timing analyses \cite{Melatos:2020rlu, Dunn:2023uqo, Meyers:2021myb, Meyers:2021slh, Kimpson:2024fel, Kimpson:2024kgq, ONeill:2024uiw, Kimpson:2025lju, 2025arXiv251011077K, Dong:2025nho}; a motivation being potential linear time computational complexity compared to a cubic one owed to Gaussian processes. State space algorithms rely on associating a system with and solving a SDE, e.g., in the Kalman filter, the prediction step (time update) is numerically solving the SDE. Analytical solutions, while limited, give clear insight, and are faster to evaluate. One possible practical use of analytical solutions in pulsar timing is to speed up the prediction step of the Kalman filter (or any filter for that matter) by replacing the numerical integration with analytic formulae. The update step remains to be carried out numerically; in this part, the mean square error between the measurement and the prediction is minimized.

Our final remark concerns the relationship between ensemble statistics and the
behaviour of a single pulsar (Section \ref{sec:averages}).  In astronomy we observe only one realization of a
stochastic process, but if the process is ergodic then time averages over a
sufficiently long observation interval converge to ensemble averages
\cite{Meyers:2021myb}. Ergodicity requires that the system explores all
accessible microstates during the observation span. Figure~\ref{fig:two_component_model_simulations}
demonstrates ergodic behaviour for the two-component neutron star model. The
trajectory of a single simulated star follows the ensemble average prediction as
time progresses. Stationary Gaussian processes such as the OU and Mat\'ern kernels
are also known to be ergodic. In this light, the results of this work can be viewed as explicitly constructing the time-domain Gaussian process equivalent of SDEs\footnote{Linear SDEs are Gaussian processes. What we meant by `explicitly constructing' is that we provide the analytic time-domain representation of the Gaussian process mean and covariance corresponding to the SDE.}; in the case of the two-component model with spin wandering, a Gaussian process realization for the angular velocities will be
\begin{equation}
    \left[ \begin{array}{c}
        \Omega_{\rm c} - \langle \Omega_{\rm c} \rangle \\ \Omega_{\rm s} - \langle \Omega_{\rm s} \rangle
    \end{array} \right] \sim
    {\cal N} \left( \left[ \begin{array}{c}
       0 \\ 0
    \end{array} \right] , \left[ \begin{array}{cc}
       \langle \Delta \Omega_{\rm c} \Delta \Omega_{\rm c} \rangle  & \langle \Delta \Omega_{\rm c} \Delta \Omega_{\rm s} \rangle  \\
        \langle \Delta \Omega_{\rm s} \Delta \Omega_{\rm c} \rangle & \langle \Delta \Omega_{\rm s} \Delta \Omega_{\rm s} \rangle 
    \end{array} \right] \right) \,,
\end{equation}
where the means and covariances are given by the analytical solutions derived in Section \ref{subsec:intrinsic_pulsar_noise}.
The dimensions can be expanded as desired to include observables such as the pulsar phase or the timing residual for PTA science purposes. This Gaussian processes route provides a direction to parameter estimation and state prediction that can complement results obtained using state space methods.

\section{Conclusions}
\label{sec:conclusions}

The techniques employed throughout this work are rooted in the classical theory
of random walks, Brownian motion, and diffusion
\cite{2010kvsp.book.....K, 1905AnP...322..549E, Langevin1908, 10.1119/1.18725, Chandrasekhar:1943ws, Uhlenbeck:1930zz, 1944BSTJ...23..282R,
1945RvMP...17..323W, 1947AmMM...54..369K}. We have applied them to problems in pulsar timing that are expressed as SDEs to draw physical insight into models that are utilized to understand observable signals.

We have obtained analytical time-domain solutions (means, covariances and PDFs) to Langevin or stochastic differential equations relevant to pulsar timing and PTAs. 
Time-domain solutions are are directly useful for physical interpretation of the dynamics in a system, but they are harder to obtain compared to their frequency-domain counterparts.
This has provided physical insights into stochastic dynamics, and we have given due attention to forces that give rise to nonstationarity in pulsar timing observables (spin frequencies and phase, or redshifts and timing residuals) \cite{1996PhRvE..54.2084G, Antonelli:2022gqw}. The degree of nonstationarity depends upon the time scales of the processes involved and the observation window. The standard practice in pulsar timing of projecting out long time deterministic trends to compute phase or timing residuals partially mitigates nonstationarity.
Specifically, we treated three models: the OU process (free Brownian particle) and the Mat\'ern-3/2 process (overdamped harmonic oscillator) for the GWB signal and pulsar red noise, deriving the Hellings--Downs correlation starting with Langevin equations; and the two-component crust--superfluid model for intrinsic pulsar timing noise, where the singular drift matrix governs the coexistence of a stationary crust-core spin lag mode and a diffusive mode (Table~\ref{tab:nonstationarity}).

Other relevant SDEs outside of the ones treated here should also be solvable analytically, such as an underdamped or critically damped Brownian harmonic oscillator \cite{2017AJ....154..220F}, a two-component model with spin wandering and memory, and a stationary two-component model with torques linear in the angular velocities. Understandably, problems, and likely the most interesting ones, cannot be dealt with analytically such as general stationary processes with steeper spectra (Appendix \ref{sec:GPs_and_SDEs}) and the two component model with magnetic dipole breaking and gravitational radiation reaction torques \cite{Meyers:2021myb,Meyers:2021slh}. 
These cases must be solved numerically. Analytical solutions, when they exist, thus serve as valuable benchmarks where the dynamics are physically transparent, i.e., tracing the observable dynamics back to the forces that generate them. In the spirit of Newton and Langevin, forces drive the dynamics of a system.

A promising direction with the results derived in this work is to integrate them with existing implementations. The Kalman filter (half of which is numerically integrating a SDE) has been applied to PTA data analysis to constrain the GWB signal \cite{Kimpson:2025lju, 2025arXiv251011077K} and to UTMOST pulsars to constrain the two-component model with spin wandering \cite{ONeill:2024uiw, Dong:2025nho}. It will be interesting to revisit radio timing noise observations and see analytical solutions leveraged to study the system \cite{Antonelli:2022gqw, ONeill:2024uiw, Antonelli:2025vhw, Dong:2025nho}. Practice also calls for scalable methods to deal with increasingly large data sets. The derivation of analytical results is halfway to producing scalable Gaussian process representations. This enables a path to parameter estimation and state prediction that would complement results obtained through state space methods.

To date, PTA analyses that utilized a state space approach have modeled the
GWB using an Ornstein-Uhlenbeck process
\cite{Kimpson:2025lju,2025arXiv251011077K}. This phenomenological choice is convenient, but it raises the question of whether a Langevin equation for the GWB can be derived beginning with first principles \cite{Liang:2024shg}. After all, the observed signal is the cumulative effect of a pulsar's electromagnetic emission that has interacted perturbatively with a large number of gravitational waves.
This line of inquiry may also provide a fresh perspective on PTA signals and data analysis through the pedagogically friendlier language of random walks.

\acknowledgements
RCB thanks Reinhard Prix for engaging discussions on the neutron star multifluid model and Wang-Wei Yu, Bruce Allen, Arian von Blanckenburg, Rutger Van Haasteren, and Colin Clark for their comments on preliminary results that have sharpened the discussion and presentation of this work. The author also grateful to Patrick Meyers for related discussions and Tom Kimpson, Nicholas O'Neill, and Andrew Melatos for important comments on an earlier version of this manuscript. The author also thanks Nicholas O'Neill and Andrew Melatos for suggesting to add the table and preparing its initial version on applications of SDEs to radio pulsar timing.

\appendix

\section{A lemma on random walks}
\label{sec:a_useful_lemma_on_random_walks}

Consider the random quantity
\begin{equation}
    X_a(t) = \int_{t_0}^t dt_1 w_a(t) \psi(t, t_1) \,,
\end{equation}
where $t$ is an observation time, $t_0$ is an initial time, $\psi(t, t_1)$ is an arbitrary function and $w_{a}(t)$ satisfies
\begin{eqnarray}
    \langle w_a(t) \rangle &=& 0 \\
    \langle w_a(t) w_b(t') \rangle &=& \Theta_{ab} \delta (t - t') \,.
\end{eqnarray}
The subscript $a$ corresponds to a set of different correlated observations of $X$; e.g., in PTA context, $a$ would be a pulsar label. The noise $w_{a}(t)$ varies extremely rapidly compared to observation and system time scales. Then, the likelihood or transition probability of realizing $X_a$ at time $t$ and $X_b$ at time $t'$ is 
\begin{align}
& \ln P(X_a, t, X_b, t'| X_{a,0}, X_{b,0}, t_0) \phantom{\dfrac{1}{1}} \nonumber \\
& = -\dfrac{1}{2} \ln | 2\pi \Theta_{ab} K(t, t', t_0) | - \dfrac{1}{2} X_a \left( \Theta_{ab} K(t, t', t_0) \right)^{-1} X_b \,, \label{eq:likelihood_XaXb}
\end{align}
where $X_{a,0}$ and $X_{b,0}$ are prior phases, and the covariance function $K(t,t',t_0)$ is given by
\begin{equation}
\label{eq:K_tt_lemma}
    K(t, t', t_0) = \int_{t_0}^t dt_1 \int_{t_0}^{t'} dt_2 \, \psi(t, t_1) \psi(t', t_2) \delta(t_1 - t_2) \,.
\end{equation}

\begin{figure}[ht]
    \centering
    \begin{tikzpicture}[scale=1.2]
        \draw[->] (0,0) -- (6,0) node[right] {$t$};
        \draw[->] (0,0) -- (0,4) node[above] {$y$};

        \draw[domain=0:5.5, smooth, variable=\x, black, line width=1pt, samples=200, opacity=0.8] plot ({\x}, {0.5 + 0.5*\x + 0.4*rand});

        \foreach \j in {1,2,3,4} {
            \pgfmathsetmacro{\xstart}{(\j-1)}
            \pgfmathsetmacro{\xend}{\j}
            \draw[dashed] (\xstart,0) -- (\xstart,{0.5 + 0.5*\xstart + 0.4*rand});
            \draw[dashed] (\xend,0) -- (\xend,{0.5 + 0.5*\xend + 0.4*rand});
            \fill[blue, opacity=0.2] (\xstart,0) rectangle (\xend,{0.5 + 0.5*\xend + 0.4*rand});
        }

        \node at (1.5,2.3) {$X_j(t)$};

        \node at (5.5,4.0) {$X(t)$};

        \draw[dashed] (0,0) -- (0,3);
        \node at (0,-0.3) {$t_0$};
        \draw[dashed] (5.5,0) -- (5.5,3);

    \end{tikzpicture}
\caption{Schematic illustration of the lemma on random walks (Appendix \ref{sec:a_useful_lemma_on_random_walks}). The random variable $X(t)$ is approximated as a sum of many independent, nonidentically distributed random steps $X_j(t)$ over small time intervals (blue rectangular bins). Because the noise varies extremely rapidly compared to observation time scales, the central limit theorem applies and $X(t)$ is Gaussian distributed.}
\label{fig:lemma_schematic}
\end{figure}
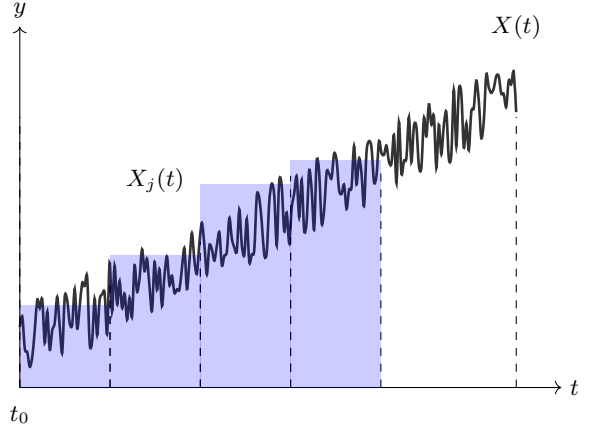

To prove this, we start with an approximation
\begin{equation}
    X_a(t) \sim \sum_{j=1}^{(t-t_0)/\Delta t} X_{a,j}(t) \,,
\end{equation}
where $X_{a,j}(t)$ are given by
\begin{equation}
    X_{a,j}(t) = \int_{t_0 + (j-1)\Delta t}^{t_0 + j \Delta t} d t_j \, w_{a}(t_j) \psi( t, t_j ) \,.
\end{equation}
For $\Delta t \rightarrow 0$, the approximation becomes exact. For $(t-t_0)/\Delta t \gg 1$, the problem of deriving the statistics of $X$ reduces to the problem of a random walk with $N_{\rm steps} \sim (t-t_0)/\Delta t$ independent and nonidentically distributed steps. The procedure is schematically illustrated in Figure \ref{fig:lemma_schematic}: the area under the noisy curve, $X(t)$, can be approximated by adding up the areas of rectangular bins, $X_j(t)$. The rapidly varying noise implies that each bin can be viewed as a independent random variable and uncorrelated with other bins. Then, the distribution of $X(t)$ can be viewed as the distribution of the sum of many independent, nonidentically distributed random steps $X_{a,j}(t)$.

For sufficiently tiny bins $\Delta t$ and $w_{a}(t)$ that varies extremely rapidly compared to observational and system time scales, we write down
\begin{equation}
    X_{a,j}(t) \sim \psi( t, \overline{t}_j ) \int_{t_0 + (j-1)\Delta t}^{t_0 + j \Delta t} d t_j \, w_{a}(t_j) \,,
\end{equation}
where $\overline{t}_j$ are the midpoints of each bin, i.e., $\overline{t}_j = t_0 + \Delta t (j - 1/2)$. The statistics of the resultant $X_a(t)$ can be determined by the statistics of the individual steps $X_{a,j}(t)$. The following can be shown;
\begin{equation}
\label{eq:Xaj_mean}
    \langle X_{a,j}(t) \rangle = 0 \,,
\end{equation}
and
\begin{equation}
\label{eq:Xaj_cov}
\begin{split}
    & \langle X_{a,j}(t) X_{b,k}(t') \rangle \phantom{\dfrac{1}{1}} \\
    & \,\,\,\,= \, \Theta_{ab} \psi(t, \overline{t}_j) \psi(t', \overline{t}_k) \phantom{\dfrac{1}{1}} \\
    & \,\,\,\,\,\,\,\,\int_{t_0+(j-1)\Delta t}^{t_0 + j \Delta t} dt_j \int_{t_0+(k-1)\Delta t}^{t_0 + k \Delta t} dt_k \, \delta(t_j - t_k) \,.
\end{split}
\end{equation}
Following the prescription that $X_a(t)$ is made up of many independent, nonidentically distributed random steps $X_{a,j}(t)$ (with a mean \eqref{eq:Xaj_mean} and covariance \eqref{eq:Xaj_cov}), then, by the central limit theorem, it follows that the random variable $X_a(t)$ is Gaussian distributed. The sum will be characterized completely by its mean and covariance. The mean vanishes,
\begin{equation}
    \langle X_a(t) \rangle \rightarrow 0 \,,
\end{equation}
because the mean of each step is zero. The covariance given by
\begin{equation}
    \langle X_a(t) X_b(t') \rangle \sim \sum_{j=1}^{(t-t_0)/\Delta t} \sum_{k=1}^{(t'-t_0)/\Delta t} \langle X_{a,j}(t) X_{b,k}(t') \rangle \,.
\end{equation}
The right hand side of the above expression can be evaluated as 
\begin{equation}
\begin{split}
    \sum_{jk} \langle X_{a,j}(t) X_{b,k}(t') \rangle \sim \Theta_{ab} K(t, t', t_0) \,.
\end{split}
\end{equation}
This defines the covariance function $K(t,t',t_0)$ given by \eqref{eq:K_tt_lemma}.
In obtaining the last expression, we have converted a double sum into a double integral. This procedure becomes increasingly valid for smaller $\Delta t \ll | t - t_0 |$.

Treating the random variable $X_a(t)$ as a sum of many independent, non-identically distributed random steps, $X_{a,j}(t)$, then the mean and covariance of $X_{a}(t)$ gives the Gaussian likelihood \eqref{eq:likelihood_XaXb}.
For $X_a = X_b = X$ and $t=t'$, the lemma reduces to the Chandrasekhar's (Lemma I in pages 23-24 of) \cite{Chandrasekhar:1943ws}; giving the PDF of the random variable $X(t)$.

We might point out that $K(t,t',t_0)$ can be written as
\begin{equation}
\label{eq:K_tt_lemma_causal_2}
    K(t, t', t_0) = \int_{t_0}^{{\rm min}(t,t')} dt_1 \, \psi(t, t_1) \psi(t', t_1) \,,
\end{equation}
where ${\rm min}(t, t') = t$ for $t < t'$ and ${\rm min}(t, t') = t'$ for $t' < t$. In addition, it can also be expressed
in a manifestly causal form;
\begin{equation}
\label{eq:K_tt_lemma_causal}
    K(t, t', t_0) = \int_{t_0}^\infty dt_1 \, \psi(t, t_1) \psi(t', t_1) \Theta(t - t_1) \Theta(t' - t_1) \,,
\end{equation}
where $\Theta(t)$ is the Heaviside step function; $\Theta(t < 0)=0$ and $\Theta( t \geq 0) = 1$.

\textit{\textbf{ Bilinear and quadratic forms.}} Throughout, we utilize the bilinear form
\begin{equation}
    V(t) W(t,t')^{-1} V(t')
\end{equation}
to express the Gaussian likelihood. In this notation, the time variables $t$ and $t'$ are viewed as continuous arguments in both the observable $V(t)$ and the kernel $W(t,t')$. The integral or sum over components is implied. In astrophysics, the quadratic form
\begin{equation}
    {\mathbf V}^{T} {\mathbf W}^{-1} {\mathbf V}
\end{equation}
with
\begin{equation}
    {\mathbf V} = \left[ \begin{array}{c}
        V(t) \\ V(t') \end{array} \right] \,
\end{equation}
and
\begin{equation}
    {\mathbf W} = \left[ \begin{array}{cc}
        W(t,t) & W(t,t') \\ W(t',t) & W(t',t') \end{array} \right] \,
\end{equation}
is commonly adopted to express the Gaussian likelihood. This is explicit for implementation and algorithmic purposes where the number of observation times is finite. The time variables $t$ and $t'$ are viewed as discrete and fixed arguments in both the observable ${\mathbf V}$ and the kernel ${\mathbf W}$.

The two notations are equivalent. The bilinear form is natural in the stochastic processes and field theory literature where quadratic forms are expressed as integrals over continuous time kernels, see e.g. \cite{Chandrasekhar:1943ws, Uhlenbeck:1930zz, 1944BSTJ...23..282R, 1945RvMP...17..323W, 1947AmMM...54..369K}. The quadratic form is the practical and explicit counterpart preferred in algorithmic contexts.

\section{Covariance of the phase in the two-component model with spin wandering}
\label{sec:cov_phase_2cm}

The covariance of the phase in the two-component model with spin wandering (Section \ref{subsec:intrinsic_pulsar_noise}) can be derived by directly using the formal solution \eqref{eq:phase_residual_formal_solution} and the covariances (\ref{eq:xip_cov}-\ref{eq:xip_xim_cross}). This gives integrals that can be evaluated analytically. The final results turn out to be
\begin{align}
    {\cal I}_1(t, t') = & \int_{t_0}^t dt_1 \int_{t_0}^{t'} dt_2 \, (t - t_1)(t' - t_2) \delta(t_1 - t_2) \nonumber \phantom{\dfrac{1}{1}}  \\
    = & -\dfrac{1}{6} ( {\rm min}(t,t')-t_0 )^2 \nonumber \phantom{\dfrac{1}{1}} \\
    & \times \left( ({\rm min}(t,t')-t_0) - 3 ({\rm max}(t,t')-t_0) \right) \,, \phantom{\dfrac{1}{1}}
\end{align}
\begin{align}
    {\cal I}_2(t,t') = & \int_{t_0}^t dt_1 \int_{t_0}^{t'} dt_2 \, \delta(t_1 - t_2) \nonumber \phantom{\dfrac{1}{1}} \\
    & \bigg[ (t-t_1)(1-e^{-(t'-t_2)/\tau}) \nonumber \phantom{\dfrac{1}{1}} \\
    & \,\,\,\,\,\,\,\,\,\,\,\,  + (t'-t_2)(1-e^{-(t-t_1)/\tau}) \bigg] \nonumber \phantom{\dfrac{1}{1}} \\
    = & - \tau^2 e^{-|t'-t|/\tau} + \tau (t - t_0 + \tau) e^{-(t'-t_0)/\tau} \nonumber \phantom{\dfrac{1}{1}} \\
    & + \tau (t' - t_0 + \tau) e^{-(t - t_0)/\tau} \nonumber \phantom{\dfrac{1}{1}} \\
    & + ({\rm min}(t,t')- t_0 - \tau)({\rm max}(t,t')-t_0 + \tau) \,, \phantom{\dfrac{1}{1}}
\end{align}
\begin{align}
    {\cal I}_3(t,t') = & \int_{t_0}^t dt_1 \int_{t_0}^{t'} dt_2 \, \delta(t_1 - t_2) \nonumber \phantom{\dfrac{1}{1}} \\
    & \,\,\,\,\,\,\,\,\,\,\,\, (1-e^{-(t-t_1)/\tau})(1-e^{-(t'-t_2)/\tau}) \nonumber \phantom{\dfrac{1}{1}}  \\
    = & - \dfrac{\tau}{2} e^{-|t'-t|/\tau} - \dfrac{\tau}{2} e^{-(t' + t - 2 t_0)/\tau} \nonumber \phantom{\dfrac{1}{1}}  \\
    & + \tau e^{-(t-t_0)/\tau} + \tau e^{-(t'-t_0)/\tau} \nonumber \phantom{\dfrac{1}{1}} \\
    & + {\rm min}(t,t')- t_0 - \tau \,. \phantom{\dfrac{1}{1}}
\end{align}
These give the covariance of the phase residual \eqref{eq:phase_residual_cov} with (\ref{eq:I1_def}-\ref{eq:I3_def}).

The large time asymptotic behavior of ${\cal I}_1$, ${\cal I}_2$, and ${\cal I}_3$ are useful to bear in mind;
\begin{align}
{\cal I}_1(t,t') & \sim \dfrac{t^2}{6}(3t'-t)\Theta(t'-t) + \dfrac{t'^2}{6}(3t-t')\Theta(t-t') \\
{\cal I}_2(t,t') & \sim -\tau^2 e^{-|t'-t|/\tau} + tt' \phantom{\dfrac{1}{1}} \\
{\cal I}_3(t,t') & \sim - \dfrac{\tau}{2} e^{-|t'-t|/\tau} + t \Theta(t'-t) + t' \Theta(t-t') \,.
\end{align}

It might be of interest that the one-component model is given by $I \dot{\Omega}=N + \xi(t)$ where $\Omega$ is an angular velocity, $I$ is a moment of intertia, and $N$ a constant torque \cite{ONeill:2024uiw, Dong:2025nho}. The noise $\xi(t)$ statistics is Gaussian with zero mean and covariance $\langle \xi(t) \xi(t') \rangle = Q I^2 \delta(t - t')$. This can be recognized to be the equation of motion of the diffusive mode $\Omega_+$. Through this mapping, the mean and covariance of the phase in the one-component model can be shown to be $\langle \phi(t) \rangle = \phi_0 + \Omega_0 (t-t_0) + N (t-t_0)^2/(2I)$ and $\langle \left( \phi(t) - \langle \phi(t) \rangle \right) \left( \phi(t') - \langle \phi(t') \rangle \right) \rangle = Q {\cal I}_1(t, t')$. This is consistent with $Q_\times = Q_- = 0$ or the vanishing stochastic torques limit of the two-component model with spin wandering. The one-component model has been used as a baseline to compare evidences with for the two-component model in \cite{ONeill:2024uiw, Dong:2025nho}.

\medskip
\section{Non-dimensionalization of the two component model}
\label{sec:nondimensionalization_of_the_two_component_model}

It is helpful to convert to non-dimensionalized variables when performing numerical simulations. In (\ref{eq:crust_eom}-\ref{eq:cross_xi_c_xi_s}), we define the dimensionless spin frequencies and time variable;
\begin{align}
    \eta &\equiv t/\tau \\
    \chi_{\rm c} &\equiv \tau \Omega_{\rm c} \\
    \chi_{\rm s} &\equiv \tau \Omega_{\rm s} \,.
\end{align}
Then, (\ref{eq:crust_eom}-\ref{eq:cross_xi_c_xi_s}) can be written as
\begin{equation}
\label{eq:two_component_state_space_form}
    \dfrac{d}{d\eta} \boldsymbol{\chi}=\boldsymbol{\Phi} \boldsymbol{\chi} + \boldsymbol{T} + \boldsymbol{W} \,,
\end{equation}
where
\begin{equation}
    \boldsymbol{\chi} = \begin{bmatrix} \chi_{\rm c} \\ \chi_{\rm s} \end{bmatrix} \,,
\end{equation}
\begin{equation}
    \boldsymbol{\Phi} = \tau \begin{bmatrix} -1/\tau_{\rm c} & 1/\tau_{\rm c} \\ 1/\tau_{\rm s} & -1/\tau_{\rm s} \end{bmatrix} \,,
\end{equation}
\begin{equation}
    \boldsymbol{T} = \tau^2 \begin{bmatrix} N_{\rm c}/I_{\rm c} \\ N_{\rm s}/I_{\rm s} \end{bmatrix} \,,
\end{equation}
and
\begin{equation}
    \boldsymbol{W} = \tau^2 \begin{bmatrix} \xi_{\rm c}/I_{\rm c} \\ \xi_{\rm s}/I_{\rm s} \end{bmatrix} \,.
\end{equation}
The noise terms satisfy
\begin{equation}
    \langle \boldsymbol{W}(\eta) \boldsymbol{W}(\eta') \rangle = \tau^3 \begin{bmatrix} Q_{\rm c} \\ Q_{\rm s} \end{bmatrix} \delta(\eta - \eta') \,.
\end{equation}
\eqref{eq:two_component_state_space_form} can be integrated using standard techniques for stochastic differential equations such as the Euler-Maruyama method \cite{1989SHH.....3..155K, 1994ZaMM...74..332V}.

\section{Gaussian processes and stochastic differential equations}
\label{sec:GPs_and_SDEs}

Random stationary Gaussian processes $y(t)$ (completely specified by a mean $\langle y(t) \rangle = 0$ and a covariance function $\langle y(t) y(t') \rangle = C(|t-t'|)$ or a one-sided PSD $S(f)$ {\it a la} Wiener-Khinchin \eqref{eq:wiener_khinchin_2PSD}) can be viewed alternatively through steady state dynamics given by a stochastic differential equation (SDE) \cite{5589113, 2013ISPM...30d..51S}. This section expresses this procedure explicitly; Section \ref{subsec:sde_to_gp} shows how to convert a SDE to a Gaussian process, and Section \ref{subsec:gp_to_sde} the other way around. Section \ref{subsec:sdes_for_gps} summarizes useful analytical conversions between Gaussian processes and SDEs.

To proceed, we setup conventions. For a time series $q(t)$, we define the Fourier transform, $\tilde{q}(\omega)$, and its inverse by the linear operations;
\begin{equation}
\label{eq:fourier_transform_def}
    \tilde{q}(\omega) = \int_{-\infty}^\infty dt \, e^{i \omega t} q(t) \,,
\end{equation}
and
\begin{equation}
\label{eq:inverse_fourier_def}
    q(t) = \dfrac{1}{2\pi} \int_{-\infty}^\infty d\omega \, e^{-i \omega t} \tilde{q}(\omega) \,,
\end{equation}
respectively. Our definition is such that positive frequency modes ($\omega = 2\pi f >0$) are associated with components $ \tilde{q}(\omega) e^{-i\omega t} $ of the time series; reminiscent of the phase $e^{- i E t / \hbar}$ that is factored in to positive energy modes in quantum mechanics.
A real function $q(t)$ would have $\tilde{q}(\omega)^* = \tilde{q}(-\omega)$. In Fourier space, the mean and covariance function of the random stationary Gaussian process $y(t)$ are $\langle \tilde{y}(f) \rangle = 0$ and $\langle \tilde{y}(f) \tilde{y}^*(f') \rangle = S(f) \delta(f-f')/2$. The modes $\tilde{y}(f)$ are decoupled in the frequency- and Fourier-domain. The one-sided PSD and the covariance function are related by the Wiener-Khinchin integral;
\begin{equation}
\label{eq:wiener_khinchin_2PSD}
    C(\tau) = \int_{0}^\infty df \, \cos(2\pi f \tau) S(f) \,.
\end{equation}

\subsection{SDE to Gaussian process}
\label{subsec:sde_to_gp}

Consider the linear time invariant SDE \eqref{eq:sde_lte} and its driving force's (white) covariance function \eqref{eq:wn_cov_sde_lte}. We first write this in state space form \cite{5589113, 2013ISPM...30d..51S}.
By defining
\begin{equation}
    {\mathbf y}(t) = \left( y(t) \ \ \dfrac{dy}{dt} \ \ \,\cdots \ \ \dfrac{d^{m-1} y}{dt} \right)^T \,,
\end{equation}
\begin{equation}
    {\mathbf F} =
    \left(
    \begin{array}{cccc}
        0 & 1 & & \\
         & \ddots & \ddots & \\
         & & 0 & 1 \\
        -a_0 & \cdots & -a_{m-2} & -a_{m-1}
    \end{array}
    \right) \,,
\end{equation}
\begin{equation}
    \mathbf{L} = \left( 0 \ \ \cdots \ \ 0 \ \ 1 \right)^T \,,
\end{equation}
then \eqref{eq:sde_lte} can be expressed as a first order vector Markov process \cite{5589113, 2013ISPM...30d..51S}:
\begin{equation}
\label{eq:sde_lte_vectorform}
    \dfrac{d}{dt} {\mathbf y}(t) = {\mathbf F} \, {\mathbf y}(t) + {\mathbf L}w(t) \,.
\end{equation}
Furthermore by defining
\begin{equation}
    {\mathbf H} = \left( 1 \ \ \cdots \ \ 0 \ \ 0 \right)^T \,
\end{equation}
we have
\begin{equation}
    y(t) = {\mathbf H} \, {\mathbf y}(t) \,. 
\end{equation}

Eq. \eqref{eq:sde_lte_vectorform} can be solved in Fourier space. Then, taking the Fourier transform of both sides of \eqref{eq:sde_lte_vectorform}, we obtain
\begin{equation}
    -i\omega \tilde{{\mathbf y}}(\omega) = {\mathbf F} \, \tilde{{\mathbf y}}(\omega) + {\mathbf L} {\tilde{w}}(\omega) \,,
\end{equation}
where $\tilde{{\mathbf y}}(\omega)$ and ${\tilde{w}}(\omega)$ are the Fourier transforms of the Gaussian process $y(t)$ and the white noise $w(t)$, respectively.
The above relation gives
\begin{equation}
    {\tilde{\mathbf y}}(\omega) = - \left( {\mathbf F} + i \omega {\mathbf I} \right)^{-1} \, {\mathbf L} \tilde{w}(\omega) \,,
\end{equation}
where ${\mathbf I}$ is the identity matrix. Therefore, the solution in Fourier space can be expressed as
\begin{equation}
\begin{split}
    \tilde{y}(\omega)={\mathbf H}\, {\tilde{\mathbf y}}(\omega) = - {\mathbf H} \, \left( {\mathbf F} + i \omega {\mathbf I} \right)^{-1} \, {\mathbf L} \tilde{w}(\omega) \,,
\end{split}
\end{equation}
and $\tilde{y}(\omega) \tilde{y}(\omega')^*$ calculated as
\begin{equation}
\begin{split}
& \tilde{y}(\omega) \tilde{y}(\omega')^* \\
&= {\mathbf H} \, \left( {\mathbf F} + i \omega {\mathbf I} \right)^{-1} \, {\mathbf L} \tilde{w}(\omega) \\
& \phantom{iiggg} \, \left[ {\mathbf H} \, \left( {\mathbf F} + i \omega' {\mathbf I} \right)^{-1} \, {\mathbf L} \tilde{w}(\omega') \right]^\dagger \\
&= {\mathbf H} \, \left( {\mathbf F} + i \omega {\mathbf I} \right)^{-1} \, {\mathbf L} \tilde{w}(\omega) \\
& \phantom{iiggg} \, \left[ \tilde{w}(\omega')^* {\mathbf L}^T \, \left[\left( {\mathbf F} - i \omega' {\mathbf I} \right)^{-1}\right]^T \, {\mathbf H}^T \right] \,.
\end{split}
\end{equation}
Noting that
\begin{equation}
    \langle \tilde{w}(\omega) \tilde{w}(\omega')^* \rangle = 2 \pi \left(\frac{s}{2}\right) \delta(\omega - \omega') \,,
\end{equation}
then we obtain the Fourier-domain covariance of $y(t)$ as
\begin{equation}
\begin{split}
& \langle \tilde{y}(\omega) \tilde{y}(\omega')^* \rangle \phantom{\dfrac{1}{1}} \\
& =  2\pi {\mathbf H} \, \left( {\mathbf F} + i \omega {\mathbf I} \right)^{-1} \, {\mathbf L} \, \left(\dfrac{s}{2}\right) \\
& \phantom{gggggii} \, \left[ {\mathbf L}^T \, \left[\left( {\mathbf F} - i \omega' {\mathbf I} \right)^{-1}\right]^T \, {\mathbf H}^T \right] \delta(\omega - \omega') \,.
\end{split}
\end{equation}
This shows that the random process $y(t)$ has a one-sided PSD given by
\begin{equation}
\label{eq:psd_sde}
\begin{split}
    S(\omega) = {\mathbf H} \, \left( {\mathbf F} + i \omega {\mathbf I} \right)^{-1} \, {\mathbf L} \, s \, {\mathbf L}^T \, \left[\left( {\mathbf F} - i \omega {\mathbf I} \right)^{-1}\right]^T \, {\mathbf H}^T \,.
\end{split}
\end{equation}
For fixed ${\mathbf F}$ determined by a SDE \eqref{eq:sde_lte}, the corresponding Gaussian process PSD can be computed using \eqref{eq:psd_sde}.

\subsection{Gaussian process to SDE}
\label{subsec:gp_to_sde}

Conversely, if the PSD of a Gaussian process $y(t)$ can be expressed in the rational form of \eqref{eq:psd_sde} \cite{5589113, 2013ISPM...30d..51S}; or
\begin{equation}
\label{eq:psd_rational}
    S(\omega) \equiv {\cal T}(i\omega) s {\cal T}(-i\omega) \,,
\end{equation}
then it can be associated with a SDE.

Another way this can also be realized is by taking the Fourier transform of \eqref{eq:sde_lte} directly. This gives
\begin{equation}
\label{eq:sde_lte_fourier}
\begin{split}
    \bigg( (-i\omega)^m & + a_{m-1}(-i\omega)^{m-1} \\
    & + \cdots + a_1 (-i\omega) + a_0 \bigg) \tilde{y}(\omega) = \tilde{w}(\omega) \,.
\end{split}
\end{equation}
Then the Fourier space solution can be made out to be
\begin{equation}
\tilde{y}(\omega) = {\cal T}(i \omega) \tilde{w}(\omega) \,,
\end{equation}
and the PSD given by \eqref{eq:psd_rational}.
However, in order to associate a SDE with a stable Markov process, the transfer functions ${\cal T}(i\omega)$ and ${\cal T}(-i\omega)$ must have all of their poles in the lower and upper half complex planes, respectively.

\begin{figure}[t]
    \centering
    \begin{tikzpicture}[scale=2]
      \draw[->] (-1.5,0) -- (1.5,0) node[right] {${\rm Re}[z]$};
      \draw[->] (0,-1.2) -- (0,1.2) node[above] {${\rm Im}[z]$};

      \draw[thick,->] (1,-0.025) -- (-1,-0.025); 
      \draw[thick,->] (-1,-0.025) arc[start angle=180,end angle=360,radius=1]; 

      \node at (-0.2,-1.1) {$C_-$};

      \node at (-0.5,-0.4) {\Large $\times$};
      \node at (0.5,-0.4) {\Large $\times$};

      \node at (0.5,-0.2) {$\mathcal{T}(z)$ poles};

      \draw[thick,->] (-1,0.025) -- (1,0.025); 
      \draw[thick,->] (1,0.025) arc[start angle=0,end angle=180,radius=1]; 

      \node at (-0.2,1.1) {$C_+$};

    \end{tikzpicture}
    \caption{Counterclockwise semicircular contours $C_-$ and $C_+$ of radius $R$ in the lower and upper half planes. The poles of $\mathcal{T}(z)$ denoted by $\times$'s are all in the lower half plane.}
    \label{fig:semicircle}
\end{figure}
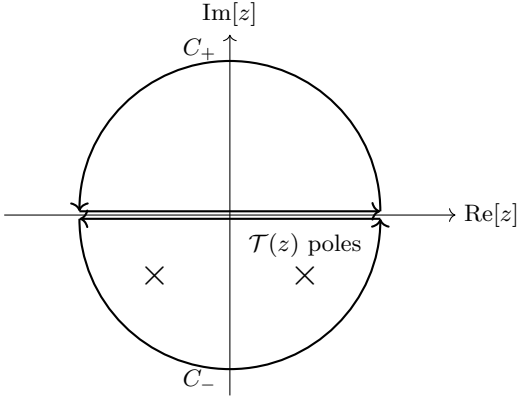

This depends on the sign convention of the Fourier transform \eqref{eq:inverse_fourier_def}; and we are using the opposite of the signal processing convention \cite{5589113, 2013ISPM...30d..51S}. In particular, the integral \eqref{eq:inverse_fourier_def} must be closed in the lower half plane for $t > 0$ (see Figure \ref{fig:semicircle});
\begin{equation}
\label{eq:inverse_fourier_def_complex}
    q(t) \equiv \dfrac{1}{2\pi} \oint_{C_-} dz \, e^{-i z t} \tilde{q}(z) \,.
\end{equation}
By the residue theorem, then
\begin{equation}
\label{eq:residue_theorem_clockwise}
    2\pi i \sum {\rm Res}\left[ \dfrac{e^{-i z t} \tilde{q}(z)}{2\pi} \right] = -\int_{-R}^R d\omega \, \dfrac{e^{-i \omega t} \tilde{q}(\omega)}{2\pi} + I(R) \,,
\end{equation}
where ${\rm Res}[f(z)]$ corresponds to the residues of a function $f(z)$, and the error term is due to the complex integral over the lower half semicircle of radius $R$:
\begin{equation}
\begin{split}
    I(R)
    &=\int dz \, \dfrac{e^{-i z t} \tilde{q}(z)}{2\pi} \bigg|_{z=Re^{i\theta}} \\
    &= \int_\pi^{2\pi} \left(d\theta \, i z(R,\theta)\right) \dfrac{e^{-i z(R,\theta) t} \tilde{q}(z(R,\theta))}{2\pi} \,.
\end{split}
\end{equation}
The phase factor $e^{-i z(R,\theta) t} = e^{-i (R \cos\theta) t} e^{- R | \sin \theta | t}$ since ${\rm Im}[z(R,\theta)] = R \sin \theta < 0$ throughout the contour. The poles of the transfer function control the temporal response of the solution; because the integral is closed in the lower half plane. In the limit $R \rightarrow \infty$, the integral $I(R) \rightarrow 0$, and we obtain
\begin{equation}
    \int_{-\infty}^\infty d\omega \, \dfrac{e^{-i \omega t} \tilde{q}(\omega)}{2\pi} = -2\pi i \sum {\rm Res}\left[ \dfrac{e^{-i z t} \tilde{q}(z)}{2\pi} \right] \,,\,\, t > 0 \,.
\end{equation}
For $t < 0$, the contour must be closed in the upper half plane (see $C_+$ in Figure \ref{fig:semicircle}). However, because all of the poles of the integral are contained in the lower half plane, the residue theorem gives
\begin{equation}
    \int_{-\infty}^\infty d\omega \, \dfrac{e^{-i \omega t} \tilde{q}(\omega)}{2\pi} = 0 \,,\,\, t < 0 \,.
\end{equation}
Combining both regions gives a stable and causal Markov process such that
\begin{equation}
    q(t) =
    \begin{cases}
        0 & \,,\,\, t < 0 \, \\
        -2\pi i \sum {\rm Res}\left[ \dfrac{e^{-i z t} \tilde{q}(z)}{2\pi} \right] & \,,\,\, t > 0 \,.        
    \end{cases}
\end{equation}

A physical way to make sense of the construction is to see that a pole $z_p = -i \omega_p$ on the negative imaginary axis will contribute a response $q(t) \sim e ^{- \omega_p t}$. This exponentially decays in time, hinting at a stable solution. In contrast, should a pole be on the positive imaginary axis, the correponsing response would be exponentially growing, or unstable. This intuition straightforwardly extends to poles away from the imaginary axis, such that poles in the lower (upper) half plane would correspond to stable (unstable) modes.

We provide an explicit construction of a SDE given a Gaussian process with the Mat\'ern-3/2 kernel. The PSD \eqref{eq:psd_matern32} is already in the desired rational form:
\begin{equation}
    S(\omega) = \dfrac{  \left( 24 \sqrt{3 }\sigma^2 / l^3 \right) }{ \left((\sqrt{3}/l) - i \omega \right)^2 \left( (\sqrt{3}/l)+i \omega \right)^2 } \,.
\end{equation}
Then by comparing to \eqref{eq:psd_rational}, we obtain the transfer function,
\begin{equation}
\label{eq:transfer_function_matern32}
    {\cal T}(i\omega) = \dfrac{1}{\left((\sqrt{3}/l) - i \omega \right)^2} \,,
\end{equation}
and the white noise PSD,
\begin{equation}
    s = \dfrac{24 \sqrt{3 }\sigma^2}{l^3} \,.
\end{equation}
Expanding the denominator of \eqref{eq:transfer_function_matern32}, we can now write down the SDE in Fourier space as
\begin{equation}
    \left( (-i\omega)^2 + \dfrac{2\sqrt{3}}{l} (-i\omega) + \left( \dfrac{\sqrt{3}}{l} \right)^2  \right) \tilde{y}(\omega) = \tilde{w}(\omega) \,,
\end{equation}
where
\begin{equation}
    \langle \tilde{w}(\omega) \rangle = 0
\end{equation}
and
\begin{equation}
    \langle \tilde{w}(\omega) \tilde{w}(\omega') \rangle = 2 \pi \left( \dfrac{ 24 \sqrt{3 }\sigma^2/{l^3} }{2} \right) \delta(\omega - \omega') \,.
\end{equation}
In the time-domain, the above SDE equations will take the form
\begin{equation}
    \left( \dfrac{d^2}{dt^2} + \dfrac{2\sqrt{3}}{l} \dfrac{d}{dt} + \left( \dfrac{\sqrt{3}}{l} \right)^2 \right) y(t) = w(t) \,,
\end{equation}
where the noise term satisfies
\begin{equation}
    \langle w(t) \rangle = 0
\end{equation}
and
\begin{equation}
    \langle w(t) w(t') \rangle = \left( \dfrac{ 24 \sqrt{3 }\sigma^2/{l^3} }{2} \right) \delta( t - t' ) \,.
\end{equation}
In state space form, the SDE becomes
\begin{equation}
    \dfrac{d}{dt} \left[ \begin{array}{c} y(t) \\ y'(t) \end{array} \right] = \left[ \begin{array}{cc}
        0 & 1  \\
        -3/l^2 & -2\sqrt{3}/l
    \end{array}  \right] \left[ \begin{array}{c} y(t) \\ y'(t) \end{array} \right] + \left[ \begin{array}{c} 0 \\ 1 \end{array} \right] w(t) \,.
\end{equation}

The conversion into a SDE can be performed exactly for Gaussian processes with the exponential kernel (OU process) and Mat\'ern kernels of arbitrary half integral order, and with approximation the radial basis function/squared exponential or Gaussian kernel \cite{5589113, 2013ISPM...30d..51S}. 
For a general process, a P\'ade approximation can be used to approximate the PSD with a rational function in the desired observation regime \cite{1992ComPh...6...82P}. The steps to obtain the corresponding SDE are the same as detailed in this section.

\subsection{SDEs for Gaussian processes}
\label{subsec:sdes_for_gps}

For the convenience of readers interested and those who want to try out their analytical prowess in deriving results (special cases), we enumerate SDEs for the commonly adopted Gaussian processes. In all cases, we list down the covariance function $C(t)$, the one-sided PSD $S(f)$, and the SDE (plus the corresponding white noise covariance); the noise term has zero mean. Readers are encouraged to consult \cite{5589113, 2013ISPM...30d..51S} for details.

\medskip
\paragraph*{\textbf{Ornstein-Uhlenbeck process (Exponential/Mat\'ern 1/2 Kernel/Free Brownian motion)}}

\begin{align}
    C(t) &= \sigma^2 e^{-|t|/l} \\
    S(f) &= \dfrac{ 4 \sigma^2 l }{ 1 + \left( 2 \pi f l \right)^2 } \\
    \left( \dfrac{d}{dt} + \dfrac{1}{l} \right) y(t) &= w(t) \\
    \langle w(t) w(t') \rangle &= \left( \dfrac{ 4\sigma^2/l }{2} \right) \delta( t - t' ) 
\end{align}



\paragraph*{\textbf{Mat\'ern-3/2 Kernel/Brownian harmonic oscillator}}

\begin{align}
    C(t) &= \sigma^2 \left( 1 + \dfrac{\sqrt{3} t}{l} \right) e^{-\sqrt{3} t/l} \\
    S(f) &= \dfrac{ 24 \sqrt{3 }\sigma^2 l }{ \left( 3 + (2\pi f l)^2 \right)^2 } \\
    \bigg( \dfrac{d^2}{dt^2} + \dfrac{2\sqrt{3}}{l} \dfrac{d}{dt} & + \left( \dfrac{\sqrt{3}}{l} \right)^2 \bigg) y(t) = w(t) \, \\
    \langle w(t) w(t') \rangle &= \left( \dfrac{ 24 \sqrt{3 }\sigma^2/{l^3} }{2} \right) \delta( t - t' ) \,
\end{align}



\paragraph*{\textbf{Radial Basis Function/Gaussian Kernel}}

\begin{align}
    \label{eq:acf_rbf} C(t) &= \sigma^2 e^{-t^2/\left(2 l^2\right)} \\
    \label{eq:psd_rbf} S(f) &= 2 \sqrt{2 \pi} \sigma^2 l e^{-2 (\pi f l)^2} \\
    \label{eq:sde_rbf}
    P_N^+\left( \dfrac{d}{dt} \right) y(t) &= w(t) \\
    \label{eq:wncov_rbf}
    \langle w(t) w(t') \rangle = & \left( \dfrac{\sqrt{2\pi} \sigma^2 2^{N+1} N! \, l^{1-2N} }{2} \right) \delta( t - t' ) \,
\end{align}
{\it $P_N^+(x)$ is obtained as follows
:} Compute roots of
\begin{equation}
    P_N(-i\omega) = \sum_{n=0}^N (-1)^n \dfrac{N! \, l^{2(n-N)}}{n! \, 2^{n-N}} \left( -i \omega \right)^{2n}
\end{equation}
for {\it even} $N$ and construct polynomials $P_N^+(x)$ and $P_N^-(x)$ such that \cite{5589113, 2013ISPM...30d..51S}
\begin{equation}
    P_N(x) = P_N^+(x) P_N^-(x) 
\end{equation}
where $P_N^+(x)$ ($P_N^-(x)$) has the roots of $P_N(x)$ with positive (negative) real parts. This construction can be obtained by writing the RBF PSD as
\begin{equation}
\begin{split}
    S \left( f(\omega) \right) &= \lim_{N\rightarrow \infty} \dfrac{ \sqrt{2\pi} \sigma^2 2^{N+1} N! \, l^{1-2N} }{ P_N( -i\omega ) } 
\end{split}
\end{equation}
and truncating the sum to finite $N$; use the identity $e^{-x} = 1/e^x = 1/\sum_{n=0}^\infty \left( x^n/n! \right)$.

\subsection{The Mat\'ern kernel}
\label{subsec:matern_kernel}

The Mat\'ern kernel with arbitrary index covers processes with ranging from the Ornstein-Uhlenbeck process (one time differentiable, shallow spectrum) to the RBF kernel (infinite differentiable, steep spectrum). This makes it a good reference point for building state space algorithms. We use this section to flesh out its state space form for arbitrary half integer index.

The Mat\'ern kernel of index $\nu > 0$ is given by
\begin{equation}
\label{eq:cov_matern_nu}
    C_\nu( t ) = \sigma^2 \dfrac{2^{1-\nu}}{\Gamma(\nu)} \left( \sqrt{2\nu} \dfrac{|t|}{l} \right)^\nu K_\nu \left( \sqrt{2\nu} \dfrac{|t|}{l} \right) \,,
\end{equation}
where $l$ is a time scale, $K_\nu$ is the modified Bessel function of the second kind, and $\Gamma(z)$ is the Gamma function. The index $\nu$ controls the smoothness of the process. The first few modified Bessel functions of the second kind are useful to note explicitly;
\begin{eqnarray}
    K_{1/2}(x) &=& \sqrt{ \dfrac{\pi x}{2} } \dfrac{e^{-x}}{x} \,, \\
    K_{3/2}(x) &=& \sqrt{ \dfrac{\pi x}{2} } \dfrac{(x+1)}{x^2} e^{-x} \,, \\
    K_{5/2}(x) &=& \sqrt{ \dfrac{\pi x}{2} } \left( \dfrac{x^2 +3x +3}{x^3} \right) e^{-x} \,.
\end{eqnarray}
In general, it is possible to write down
\begin{equation}
    K_{p + 1/2}(x) = \sqrt{\dfrac{\pi x}{2}} e^{-x} f_p(x) \,,
\end{equation}
where $f_p(x)$ satisfies the recursion relation
\begin{equation}
    f_p(x) = f_{p-2}(x) + (2p-1) \dfrac{f_{p-1}(x)}{x} \,,
\end{equation}
together with $f_0(x)=1/x$ and $f_1(x)=(x+1)/x^2$. See pages 443-445 of Abramowitz and Stegun 1972 \cite{1972hmfw.book.....A}. The above formulae can be used to show that $C_{1/2}(t)$ and $C_{3/2}(t)$ reduce exactly to the exponential and Mat\'ern 3/2 kernel expressions given in Section \ref{subsec:sdes_for_gps}.

The PSD of the covariance function \eqref{eq:cov_matern_nu} is given by
\begin{equation}
\label{eq:psd_matern_nu}
\begin{split}
    S_\nu(f) = \, & 4 \sigma^2 \sqrt{\pi} \dfrac{ \Gamma(\nu + 1/2)}{\Gamma(\nu)} \left( \dfrac{\sqrt{2\nu}}{l} \right)^{2\nu} \\
    & \times \left( \left( \dfrac{\sqrt{2\nu}}{l} \right)^2 + (2\pi f)^2 \right)^{-(\nu + 1/2)} \,.
\end{split}
\end{equation}
Substituting \eqref{eq:psd_matern_nu} into the Wiener-Khinchin integral \eqref{eq:wiener_khinchin_2PSD} recovers \eqref{eq:cov_matern_nu}. Equation 8.432.5 of TISP7 (page 917) \cite{2007tisp.book.....G} gives
\begin{equation}
    K_\nu(xz)= \dfrac{\Gamma(\nu+1/2)}{\Gamma(1/2)} \left( \dfrac{2z}{x} \right)^\nu \int_0^\infty dt \, \dfrac{\cos(xt)}{(z^2 + t^2)^{\nu + 1/2}} \,,
\end{equation}
for ${\rm Re}(\nu+1/2)>0$, $x > 0$, and ${\rm arg} z < \pi/2$; from which the covariance function \eqref{eq:cov_matern_nu} can be verified by performing \eqref{eq:wiener_khinchin_2PSD}. 

The SDE can be written for half-integer values of $\nu = p + 1/2$ with $p = 0, 1, 2, \ldots$. This is particularly straightforward since the PSD \eqref{eq:psd_matern_nu} is in the desired rational form. In this case, we can rewrite the PSD as \eqref{eq:psd_rational} with
\begin{equation}
\label{eq:transfer_func_matern_nu}
{\cal T}(i\omega) = \left( \dfrac{\sqrt{2p+1}}{l} - i \omega \right)^{-(p+1)}
\end{equation}
and
\begin{equation}
\label{eq:wnpsd_matern_nu}
s=4 \sigma^2 \sqrt{\pi} \dfrac{ \Gamma(p+1)}{\Gamma(p+1/2)} \left( \dfrac{\sqrt{2p+1}}{l} \right)^{2p+1} \,,
\end{equation}
where $S_{p+1/2}(\omega)=S_{p+1/2}(f=2\pi\omega)$. The above results can be compared with the special cases of the exponential ($p=0$) and the Mat\'ern ($p=1$) kernels in Section \ref{subsec:sdes_for_gps}; recall that the coefficients of the denominator of the transfer function when expanded appear as the coefficients of the SDE. In general, the SDE for a Mat\'ern kernel with half integral order ($\nu=p+ 1/2$) will have the operator form
\begin{equation}
\label{eq:sde_matern_nu}
    \left( \dfrac{d}{dt} + \dfrac{\sqrt{2p+1}}{l} \right)^{p+1} y(t) = w(t) \,,
\end{equation}
where the noise covariance is given by
\begin{equation}
\label{eq:wncov_matern_nu}
    \langle w(t) w(t') \rangle = \dfrac{ s }{2} \delta( t - t' ) \,
\end{equation}
where $s$ is given by \eqref{eq:wnpsd_matern_nu}. The operator appearing in \eqref{eq:sde_matern_nu} should be understood as a polynomial of order $p+1$; i.e., using the binomial expansion,
\begin{equation}
\begin{split}
    & \left( \dfrac{d}{dt} + \dfrac{\sqrt{2p+1}}{l} \right)^{p+1} \\
    & \,\,\,\, \equiv \sum_{k=0}^{p+1} \binom{p+1}{k} \left( \dfrac{\sqrt{2p+1}}{l} \right)^{p+1-k} \dfrac{d^k}{dt^k} \,,
\end{split}
\end{equation}
where the binomial coefficient is given by 
\begin{equation}
    \binom{p+1}{k} = \dfrac{(p+1)!}{k! (p+1-k)!} \,.
\end{equation}
The coefficient of the highest derivative $d^{p+1}/dt^{p+1}$ is unity. In the form of \eqref{eq:sde_matern_nu} together with \eqref{eq:wncov_matern_nu}, the SDE will be of $(p+1)$-th order. The state will be of dimension $p+1$; since $p+1$ initial conditions will be required to fully specify the system.

To end, we write down the transition matrix ${\mathbf F}$ in the state-space representation of the SDE \eqref{eq:sde_matern_nu}; see Section \ref{subsec:sde_to_gp}. The matrix ${\mathbf F}$ is given by
\begin{equation}
    {\mathbf F} =
    \left(
    \begin{array}{cccc}
        0 & 1 & & \\
         & \ddots & \ddots & \\
         & & 0 & 1 \\
        -a_0 & \cdots & -a_{p-1} & -a_{p}
    \end{array}
    \right) \,,
\end{equation}
where the coefficients $a_k$ are given by
\begin{equation}
    a_k = \binom{p+1}{k} \left( \dfrac{\sqrt{2p+1}}{l} \right)^{p+1-k} \,,
\end{equation}
for $k = 0, 1, \ldots, p$.

\section{Generating Mat\'ern processes}
\label{sec:generating_matern}

It is not a new result that the integral of the OU process is nonstationary \cite{1996PhRvE..54.2084G, Antonelli:2022gqw}. Physically, this is simply the mean square displacement of a free Brownian particle that grows linearly in observation time \cite{Uhlenbeck:1930zz,Chandrasekhar:1943ws}.
But it is also a good question to ask how this view is consistent with the SDE realizations of Mat\'ern processes; where the OU process can be seen as a special case.

A differential operator acting on the Mat\'ern 3/2 process would result in the OU process. To see this, consider an OU process $z_{1/2}$ that satisfies the operation
\begin{equation}
(D + \lambda) z_{1/2} = w
\end{equation}
where $D$ is a short for the differential operator $d/dt$ and $w$ is white noise. The PSD of $z_{1/2}$ would then be
\begin{equation} S_{1/2} = \dfrac{|\tilde{w}|^2}{ (\omega^2 + \lambda^2) } \,. \end{equation}
To derive this and the following PSDs, one only has to take the Fourier transform of both sides of the equation and multiply the result with its complex conjugate.
The Mat\'ern 3/2 process $z_{3/2}$ is related to the OU process by
\begin{equation} (D + \lambda) z_{3/2} = z_{1/2} \,. \end{equation}
Then, its PSD can be shown to be
\begin{equation} S_{3/2} = \dfrac{ S_{1/2} }{ (\omega^2 + \lambda^2) } = \dfrac{ |\tilde{w}|^2 }{ (\omega^2 + \lambda^2)^2 } \,. \end{equation}
The inverse operator $(D + \lambda)^{-1}$ acting on the OU process $z_{1/2}$ therefore produces a stationary Gaussian process, that is
\begin{equation} z_{3/2} = (D + \lambda)^{-1} z_{1/2} \,. \end{equation}
The logic extends straightforwardly to construct higher order Mat\'ern processes, $z_{n/2}$ for odd $n=1,3,5,$ and so on. Generally, 
\begin{equation} z_{n/2} = [(D + \lambda)^{-1}]^{(n-1)/2} z_{1/2} \,, \end{equation}
where $n$ is an odd, positive integer. The operator $(D + \lambda)^{-1}$ acting $(n-1)/2$ times on the OU process $z_{1/2}$ produces a stationary process with a PSD that is infrared convergent (or goes to a constant as $\omega \rightarrow 0$).
The PSD can be shown to be
\begin{equation}
    S_{n/2}=\dfrac{ |\tilde{w}|^2 }{ (\omega^2 + \lambda^2)^{(n+1)/2} } \,.
\end{equation}
This gives $S_{n/2}\sim |\tilde{w}|^2/\lambda^{n+1}$ in the zero frequency limit.

In contrast, the integral operation $\int dt$ on $z_{1/2}$ produces a nonstationary process, $\int dt \, z_{1/2}$, with an effective PSD $\sim S_{1/2}/\omega^2$ that is infrared divergent, characteristic of nonstationarity.


%

\end{document}